\documentclass[a4paper]{article}

\usepackage{INTERSPEECH2021}

\usepackage{cite}
\usepackage{amsmath,amssymb,amsfonts}
\usepackage{algorithmic}
\usepackage{graphicx}
\usepackage{textcomp}
\usepackage{multirow}
\usepackage{hrefhide}
\usepackage{mathtools}
\usepackage{subcaption}
\DeclarePairedDelimiter{\ceil}{\lceil}{\rceil}

\newcommand*{\dittoclosing}{ \raisebox{-0.5ex}{''} }

\usepackage{color, colortbl}
\definecolor{LightCyan}{rgb}{0.88,1,1}

\usepackage[first=0,last=9]{lcg}

\begin{document}
\title{Automatic non-invasive Cough Detection based on Accelerometer and Audio Signals}
\name{Madhurananda Pahar, Igor Miranda, Andreas Diacon, Thomas Niesler}
\address{Department of Electrical and Electronic Engineering, Stellenbosch University, South Africa }
\email{\{mpahar, trn\}@sun.ac.za}

\maketitle

\begin{abstract}

	We present an automatic non-invasive way of detecting cough events based on both accelerometer and audio signals. 
	The acceleration signals are captured by a smartphone firmly attached to the patient's bed, using its integrated accelerometer. 
	The audio signals are captured simultaneously by the same smartphone using an external microphone. 
	We have compiled a manually-annotated dataset containing such simultaneously-captured acceleration and audio signals for approximately 6000 cough and 68000 non-cough events from 14 adult male patients in a tuberculosis clinic. 
	LR, SVM and MLP are evaluated as baseline classifiers and compared with deep architectures such as CNN, LSTM, and Resnet50 using a leave-one-out cross-validation scheme. 
	We find that the studied classifiers can use either acceleration or audio signals to distinguish between coughing and other activities including sneezing, throat-clearing, and movement on the bed with high accuracy. 
	However, in all cases, the deep neural networks outperform the shallow classifiers by a clear margin and the Resnet50 offers the best performance by achieving an AUC exceeding 0.98 and 0.99 for acceleration and audio signals respectively. 
	While audio-based classification consistently offers a better performance than acceleration-based classification, we observe that the difference is very small for the best systems. 
	Since the acceleration signal requires less processing power, and since the need to record audio is sidestepped and thus privacy is inherently secured, and since the recording device is attached to the bed and not worn, an accelerometer-based highly accurate non-invasive cough detector may represent a more convenient and readily accepted method in long-term cough monitoring. 
	
\end{abstract}


\section{Introduction}

Coughing is a common symptom of respiratory disease and the forceful expulsion of air to clear up the airway \cite{korpavs1996analysis}. 
It is distinctive in nature than the other audio events and is an important indicator used by physicians for clinical diagnosis and health monitoring in more than 100 respiratory diseases \cite{knocikova2008wavelet}, including tuberculosis (TB) \cite{botha2018detection}, asthma \cite{al2013signal}, pertussis \cite{pramono2016cough} and COVID-19 \cite{carfi2020persistent}. 
Machine learning algorithms can be applied on the acoustic features extracted from the cough audio for automatic cough detection and classification \cite{miranda2019comparative, laguarta2020covid, pahar2020covid}. 
However, using an audio-based monitoring system creates privacy issues \cite{chen2008audio}, specially when the audio is captured by a smartphone \cite{tung2019exploiting, xia2020pams} and unique filtering processes might be required to preserve the privacy for continuous monitoring \cite{liaqat2017method}.

Thus, using accelerometers can be an alternative to the audio. 
Moreover, due to its much lower sampling rates, it requires less computing and processing power than the audio \cite{mehigan2009harnessing}. 
Automatic cough detection based on accelerometer measurements is also possible when the device is placed on the patient's body and the acceleration signals are used for feature extraction \cite{mohammadi2019automatic}.
An accelerometer is insensitive to environmental and background noise, thus it can be used in conjunction with other sensors such as microphones, ECG and thermistors \cite{munyard1994new}. 
Body-attached accelerometers have for example proved to be useful in detecting coughs when placed in contact with a patient's throat~\cite{coyle2010systems, fan2014cough} or at the laryngeal prominence (Adam's apple)~\cite{mohammadi2019automatic}. 
A cough monitoring system using a contact microphone and an accelerometer attached to the participant's suprasternal (jugular) notch was developed in~\cite{pavesi2001application}.
The participants moved around their homes while the cough audio and vibration was recorded. 
A similar ambulatory cough monitoring system, using an accelerometer attached to the skin of the participant's suprasternal notch using a bioclusive transparent dressing, was developed in~\cite{paul2006evaluation}.
Here, the recorded signal is transmitted to a receiver carried in a pocket or attached to a belt.  
Two accelerometers, one placed on the abdomen and the second on a belt wrapped at dorsal region, have been used to measure cough rate after cross-correlation of the two sensor signals~\cite{chan2014systems}. 
Regression analysis, carried out on both audio and accelerometer signals gathered from 50 children,
was able to achieve 97.8\% specificity and 98.8\% sensitivity when the accelerometer was placed in the centre of the abdomen between the navel and sternal notch~\cite{hirai2015new}. 
Finally, multiple sensors, including ECG, thermistor, chest belt, accelerometer and audio microphones were used for cough detection in~\cite{drugman2013objective}. 

However, attaching an accelerometer to the patient's body is inconvenient and intrusive.
We propose the monitoring of coughing based on the signals captured by the on-board accelerometer of an inexpensive consumer smartphone firmly attached to the patient's bed, as shown in Figure \ref{fig:recorder-position}.
This eliminates the need to wear a measuring equipment.
The work presented here extends our previous study \cite{pahar2021deep} by using three additional classifiers in the cough detection process and by comparing the performance between the proposed accelerometer-based classifiers and the baseline systems that classify audio signals of the same cough events. 
Such audio-based cough detection systems have been reported to discriminate between coughing and other sounds with areas under the ROC curve (AUCs) as high as 0.96 ~\cite{miranda2019comparative} and specificity as high as 99\% \cite{amoh2015deepcough}.
Although we have found that the audio-based cough detection still outperforms the accelerometer-based detection, we demonstrate that the difference in the performance is narrow, as
our best 50-layer residual architecture (Resnet50) based cough detector achieves an AUC of 0.996  
for audio-based and 0.989 
for accelerometer-based detection respectively. 
Thus we present an automatic non-invasive accelerometer-based cough detection system which can be used in long-term monitoring of patient's recovery process. 

The structure of the reminder of this paper is as follows.
Section \ref{sec:data} describes data collection while Section \ref{sec:feat} details the features we extract from this data. 
The classifiers we use for experimentation are explained in Section~\ref{sec:classifier} and the classification process itself is elaborated in Section~\ref{sec:classification}.
The results are presented in Section \ref{sec:results} and discussed in Section~\ref{sec:discussion}.
Finally, Section \ref{sec:conclusion} concludes the paper.

\section{Dataset Preparation}\label{sec:data}

\subsection{Data collection}

Data has been collected at a small 24h TB clinic near Cape Town, South Africa, which can accommodate approximately 10 staff and 30 patients. 
The clinic contains 8 wards and each ward has four beds, thus four patients at one time can be monitored inside a ward. 
The overall motivation of this study was to develop a practical method of automatic cough monitoring for the patients in this clinic, so that the progress of the recovery process can be monitored.  

\begin{figure}[h!]
	\centerline{\includegraphics[width=0.5\textwidth]{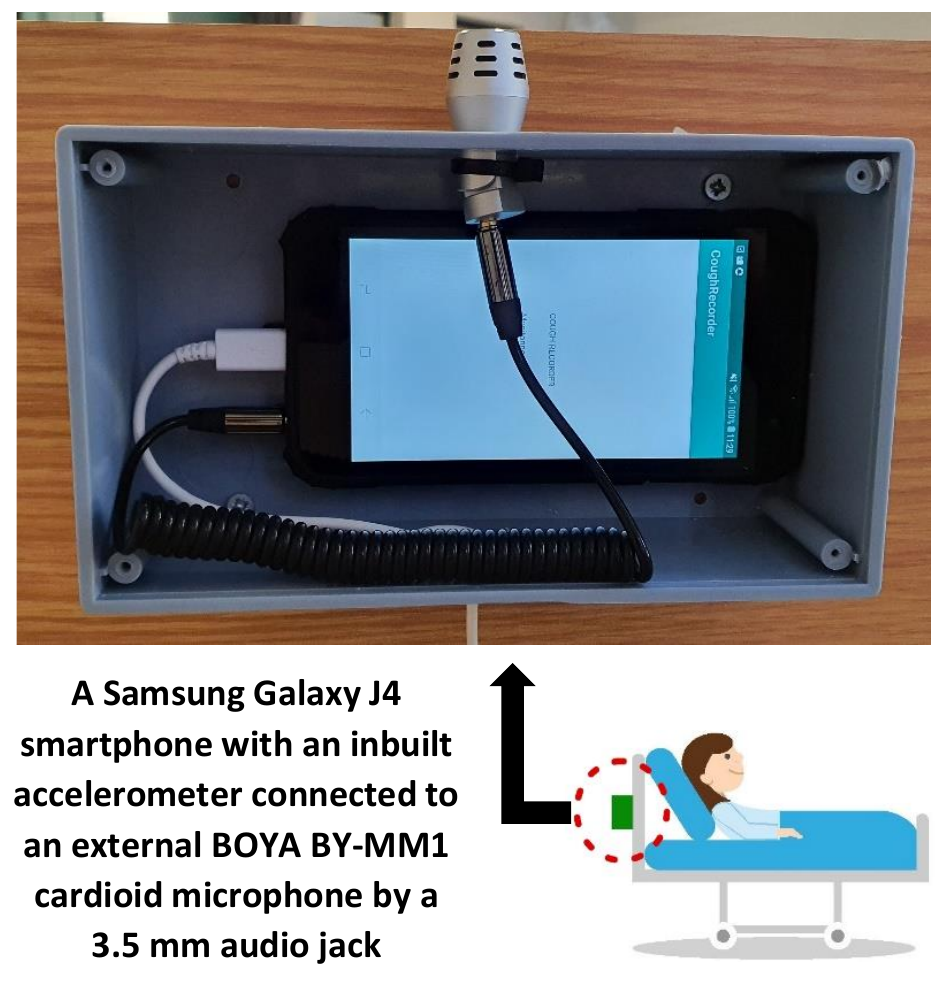}}
	\caption{\textbf{Recording Equipment:} A plastic enclosure housing an inexpensive smartphone running data gathering software is attached behind the headboard of each bed. The acceleration signal from the on-board accelerometer as well as the audio signal from the external microphone are monitored. Recording is triggered if activity is detected in either of these two signals. }
	\label{fig:recorder-position}
\end{figure}

Figure \ref{fig:recorder-position} shows the recording setup, where 
an enclosure housing an inexpensive consumer smartphone is firmly attached to the back of the headboard of each bed in a ward. 
An Android application, developed specifically for this study, monitors the accelerometer and the audio signals.
The on-board smartphone accelerometer has a sampling frequency of $100$Hz.
Although this sensor provides tri-axial measurements, we record only the vector magnitude.
A BOYA BY-MM1 external microphone was used to capture audio signals (visible in Figure~\ref{fig:recorder-position}) and it has the sampling rate of 22.05 kHz. 
Using a simple energy detector, activity on either the acceleration or the audio channels triggers the simultaneous recording of both.
This results in a dataset consisting of a sequence of non-overlapping tome intervals during which both acceleration and audio have been recorded.

\subsection{Data annotation}

A large volume of both the audio and accelerometer data have been captured by using this energy-threshold-based detection for both audio and acceleration signals. 
Ceiling-mounted cameras recorded continuous video to assist with the data annotation process. 
The audio signals and the video recordings allowed the presence or absence of a cough in an event to be unambiguously confirmed during manual annotation.
In the remainder of this paper, we will define an `event' to be any interval of activity in either the accelerometer or the audio signals. 

\begin{figure}[!h]
	\centerline{\includegraphics[width=0.5\textwidth]{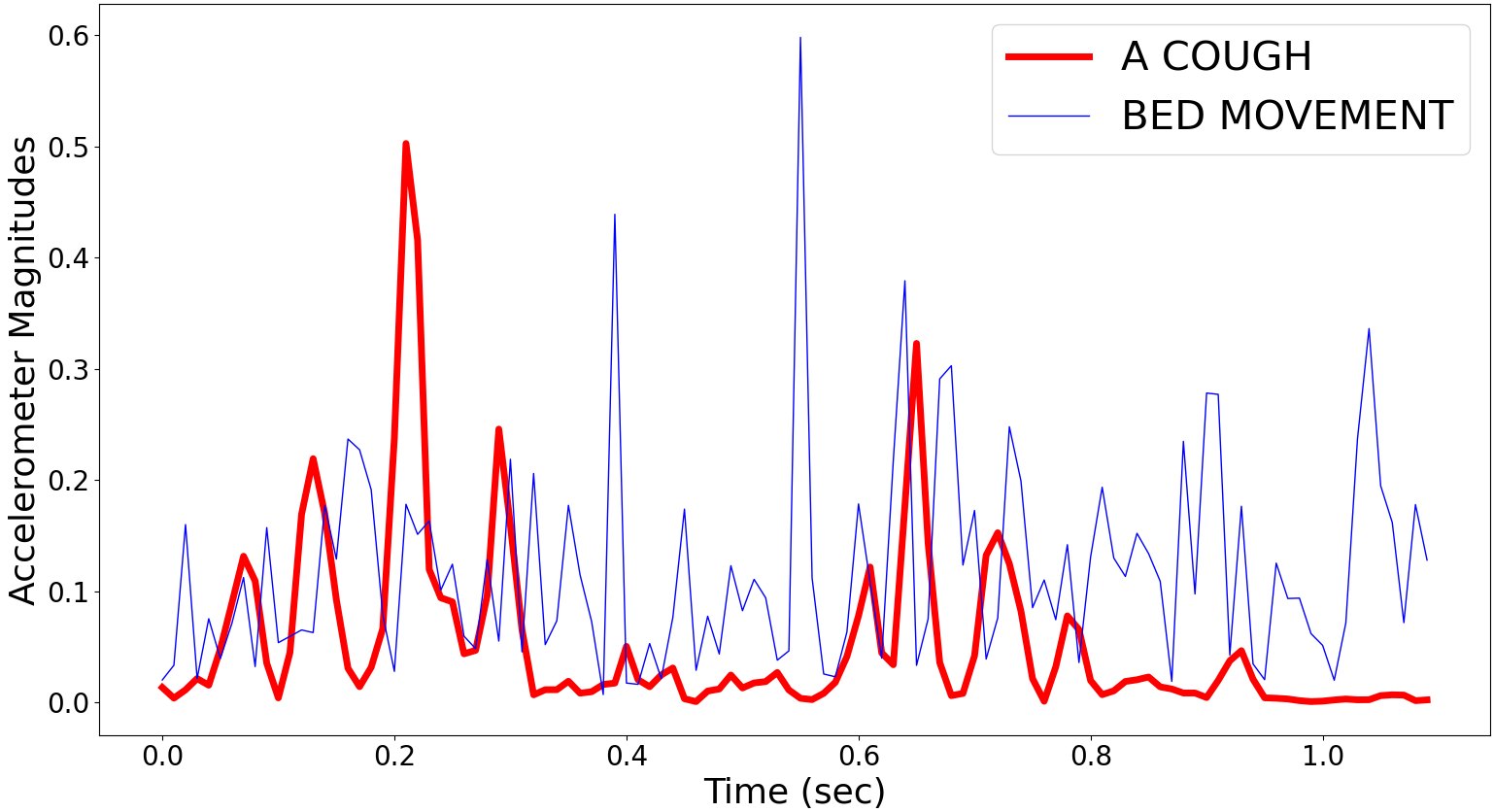}}
	\caption{\textbf{The accelerometer magnitudes} for a cough event (red) and a non-cough event (blue). In this case, the non-cough event was the patient moving on the bed. 
	}
	\label{fig:cough-Ncough-acc}
\end{figure}

\begin{figure*}
	\centering
	
	\begin{subfigure}{.49\textwidth}
		\centering
		\includegraphics[width=\linewidth]{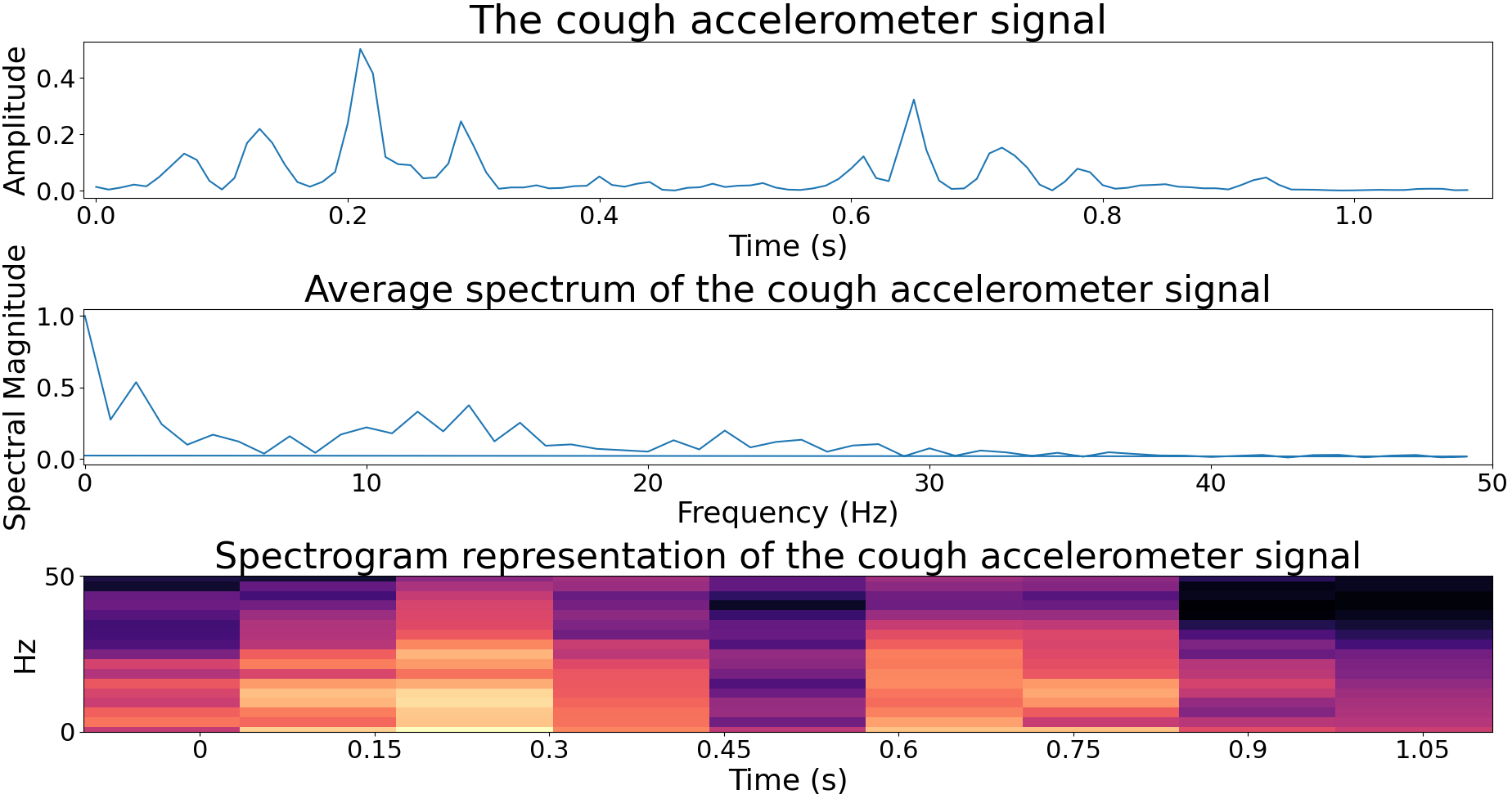}
		\caption{Accelerometer signal of the cough event. }
		\label{fig:sub1}
	\end{subfigure}%
	\hspace{5pt}
	\begin{subfigure}{.49\textwidth}
		\centering
		\includegraphics[width=\linewidth]{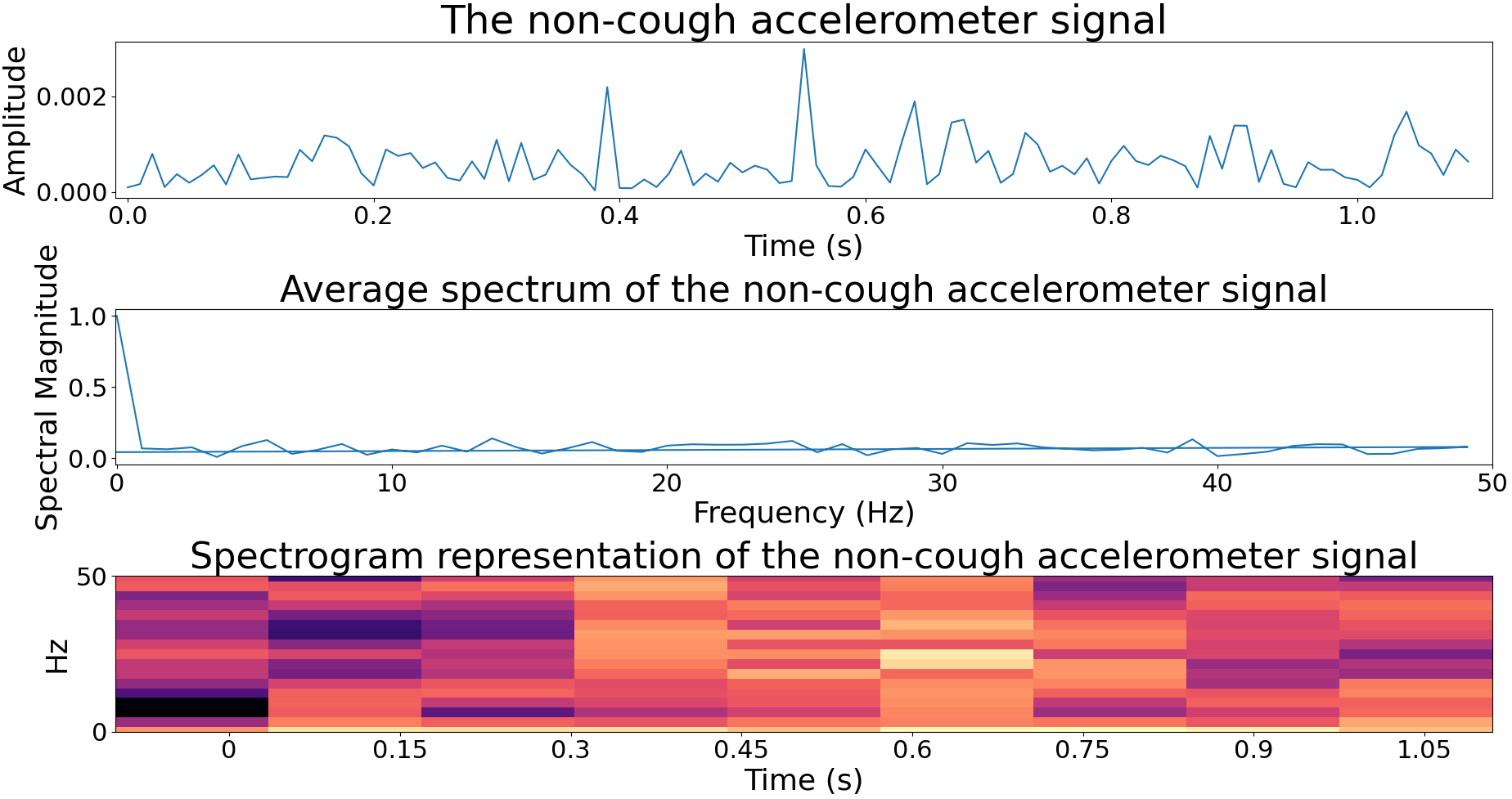}
		\caption{Accelerometer signal of the non-cough (bed-movement) event. }
		\label{fig:sub2}
	\end{subfigure}
	
	\vspace{15pt}
	
	\begin{subfigure}{.49\textwidth}
		\centering
		\includegraphics[width=\linewidth]{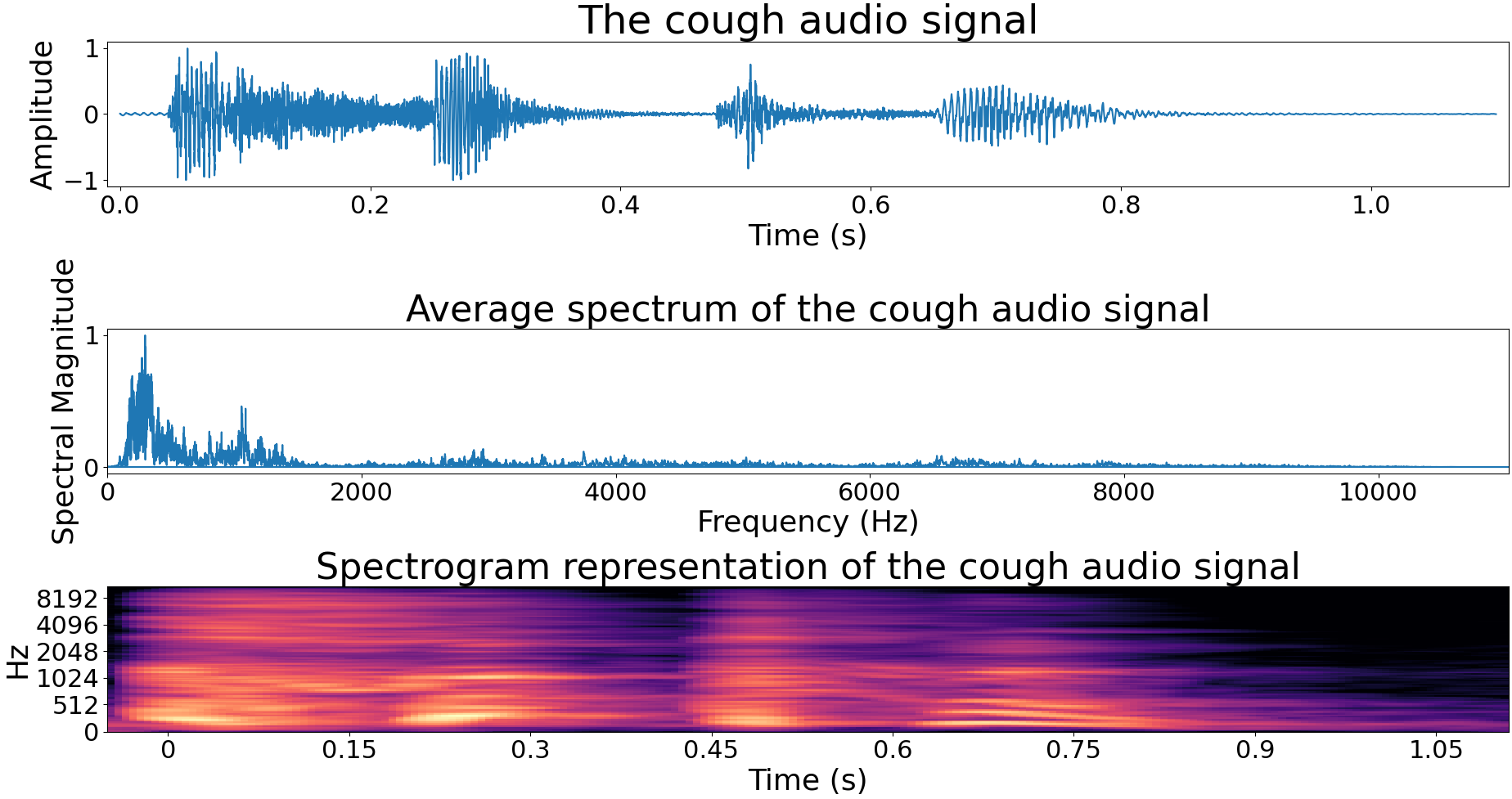}
		\caption{Audio signal of the cough event. }
		\label{fig:sub3}
	\end{subfigure}%
	\hspace{5pt}
	\begin{subfigure}{.49\textwidth}
		\centering
		\includegraphics[width=\linewidth]{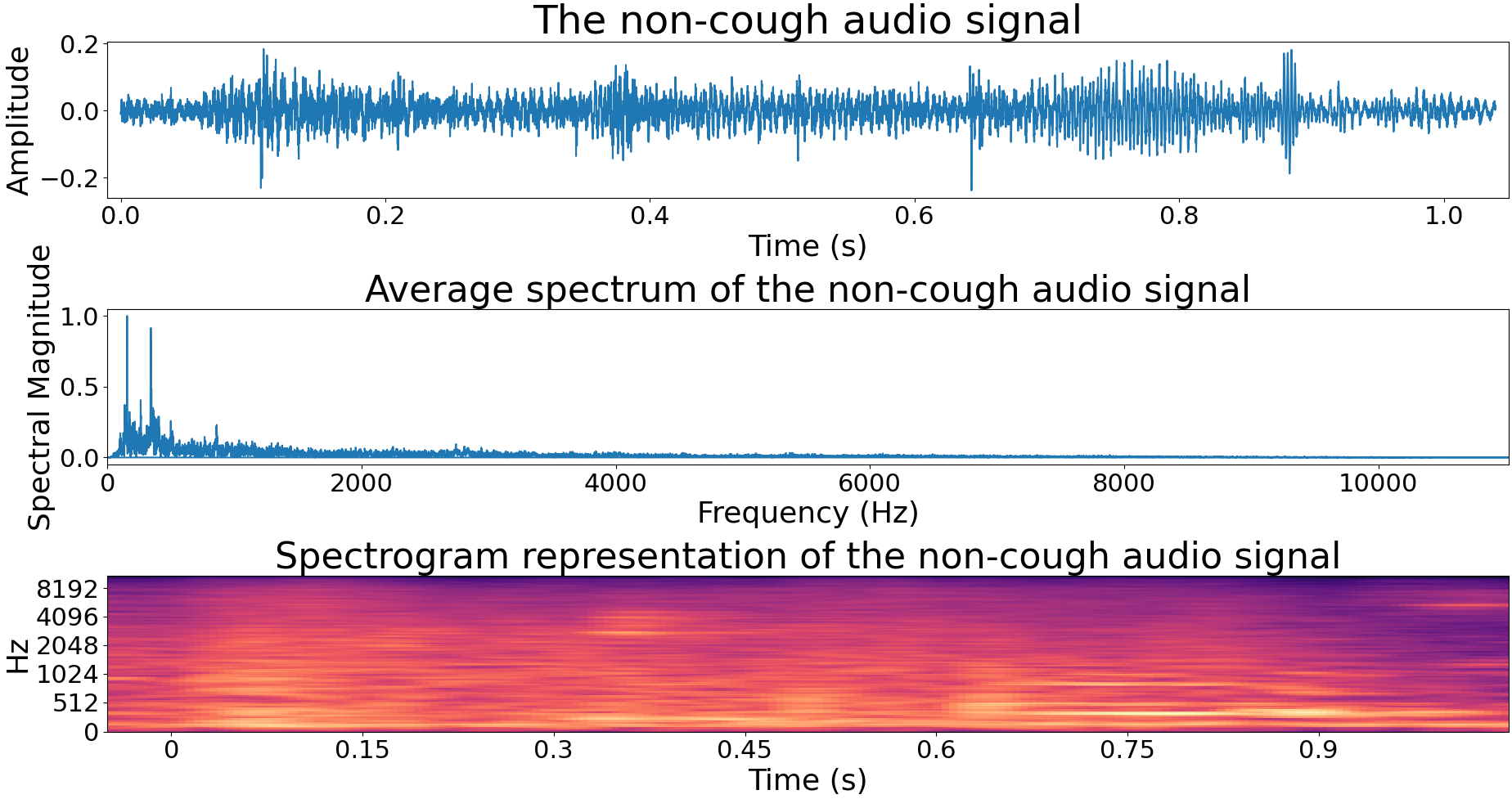}
		\caption{Audio signal of the non-cough (bed-movement) event. }
		\label{fig:sub4}
	\end{subfigure}
	
	\caption{\textbf{Spectrogram representation of the cough and non-cough events shown in Figure~\ref{fig:cough-Ncough-acc}:} The cough event is shown on the left and the non-cough event (bed-movement) on the right. The accelerometer and audio signals are shown at the top and bottom row respectively. The audio signal has a higher sampling rate and thus contains more frequency and time-domain information than accelerometer measurements. }
	\label{fig:cough-compare}
\end{figure*}

The non-cough events are generated mostly due to the patients getting in and out of the bed, moving while on the bed, sneezing and throat-clearing. 
Examples of the accelerometer magnitude signals for a cough event and a non-cough event (in this case moving on the bed) are shown in Figure \ref{fig:cough-Ncough-acc}. 
The spectrogram representations of these two signals are shown in Figure~\ref{fig:cough-compare}. 
Manual annotation was performed using the multimedia software tool ELAN \cite{wittenburg2006elan}, which allowed easy consolidation of the accelerometer, audio and video signals for accurate manual labelling. 

\subsection{Final Dataset}

The final dataset, summarised in Table~\ref{table:Ground-Truth-Dataset-Summary}, contains approximately 6000 coughs and 68000 non-coughs events from 14 adult male patients. 
Cough events are on average  1.90 sec 
long, with a standard deviation of 0.26 sec. 
Non-cough events are on average 1.70 sec 
long, with a standard deviation of 0.24 sec. 
The total lengths of the cough and non-cough events are 11397.60 sec (3.16 hours) and 115928.12 sec (32.20 hours) respectively. 
No other information regarding patients are recorded due to the ethical constraints of the study. 
This dataset was used to train and evaluate six classifiers, explained in Section~\ref{sec:classifier} within a leave-one-out cross-validation framework, described in Section~\ref{sec:classification}.

\begin{table}[h!]
	\footnotesize
	\renewcommand\arraystretch{1.4}
	\setlength\minrowclearance{1.0pt}
	\setlength{\tabcolsep}{4pt} 
	\caption{\textbf{Ground Truth Dataset Summary:} `\uppercase{Patients}': list of the patients; `\uppercase{Coughs}': number of confirmed cough events; `\uppercase{Non Coughs}': number of confirmed events that are not coughs; `\uppercase{Cough time}': total amount of time (in sec) for cough events; `\uppercase{Non-cough time}': total amount of time (in sec) for non-cough events. } 
	\centering 
	\begin{center}
		\begin{tabular}{ c c c c c }
			\hline
			\hline
			{\multirow{2}{*}{\textbf{\uppercase{Patients}}}} & {\multirow{2}{*}{\textbf{\uppercase{Coughs}}}} & \textbf{\uppercase{Non}} & \textbf{\uppercase{Cough}} & \textbf{\uppercase{Non-}} \\
			&  & \textbf{\uppercase{coughs}} & \textbf{\uppercase{time}} & \textbf{\uppercase{cough time}} \\
			\hline 
			\hline
			Patient 1 & 88 & 973 & 169.16 & 1660.67 \\
			
			\hline
			Patient 2 & 63 & 1111 & 117.67 & 1891.92 \\
			
			\hline
			Patient 3 & 469 & 11025 & 893.91 & 18797.32 \\
			
			\hline
			Patient 4 & 109 & 9151 & 204.06 & 15596.71 \\
			
			\hline
			Patient 5 & 97 & 7826 & 188.26 & 13344.98 \\
			
			\hline
			Patient 6 & 192 & 12437 & 360.72 & 21197.35 \\
			
			\hline
			Patient 7 & 436 & 14053 & 825.23 & 23953.15 \\
			
			\hline
			Patient 8 & 368 & 2977 & 702.05 & 5077.89 \\
			
			\hline
			Patient 9 & 2816 & 3856 & 5345.27 & 6569.32 \\
			
			\hline
			Patient 10 & 649 & 2579 & 1236.84 & 4400.42 \\
			
			\hline
			Patient 11 & 205 & 527 & 391.42 & 901.38 \\
			
			\hline
			Patient 12 & 213 & 323 & 402.61 & 547.62 \\
			
			\hline
			Patient 13 & 213 & 712 & 401.61 & 1211.75 \\
			
			\hline
			Patient 14 & 82 & 455 & 158.77 & 777.64 \\
			
			\hline
			\textbf{TOTAL} & \textbf{6000} & \textbf{68005} & \textbf{11397.6} & \textbf{115928.12} \\
			
			\hline
			\hline
		\end{tabular}
	\end{center}
	\label{table:Ground-Truth-Dataset-Summary}
\end{table}

\subsection{Dataset balancing}

According to Table \ref{table:Ground-Truth-Dataset-Summary}, cough events are outnumbered by the non-cough events in our dataset. 
This imbalance can affect the machine learning classifiers detrimentally \cite{van2007experimental,krawczyk2016learning}.
So, we have applied synthetic minority oversampling technique (SMOTE) to create new synthetic samples of the minor class instead of for example oversampling randomly \cite{chawla2002smote,lemaitre2017imbalanced} while training the classifiers. 
This way we have addressed this class imbalance for both the accelerometer and audio events. 
SMOTE has previously been successfully applied to cough detection and classification based on audio recordings \cite{windmon2018tussiswatch, pahar2020covid, pahar2021deep}.

\section{Feature Extraction}\label{sec:feat}

The feature extraction process is illustrated for both accelerometer and audio signal in Figure \ref{fig:feat-extract}. 

\subsection{Accelerometer features}

Power spectrum, root mean square (RMS) value, kurtosis, moving averages and crest factor are extracted from the accelerometer magnitude samples. 
No de-noising has been applied prior to the feature extraction process. 
Power spectra~\cite{bingham1967modern} has been used to represent sensor data for input to classifiers, including neural networks, in several studies \cite{jung1997estimating,durak2003short,sinha2003artificial, liang2013application}. 
RMS \cite{levinson1947wiener} values from the sensor data have also been found to be useful features \cite{lux2014application, gilmore2017assessing}. 
The kurtosis has also been useful for machine learning applications as it indicates the prevalence of higher amplitudes~\cite{zhang2013machine}. 
Moving averages indicate the smoothed evolution of a signal over a time period and have been found to be useful features for sensor analysis \cite{yuan2017wind}.
Finally, the crest factor measures the ratio of the peak and the RMS signal amplitudes and have also been found to help machine learning prediction \cite{lepine2017use} including deep learning \cite{ren2017multi}.

\subsection{Audio features}

Features such as mel-frequency cepstral coefficients (MFCCs), zero crossing rate (ZCR) and kurtosis are extracted from the audio signal. 
MFCCs are successfully used as features in audio analysis and especially in automatic speech recognition \cite{WeiHan2006, pahar_coding_2020}. 
They can differentiate dry coughs from wet coughs \cite{chatrzarrin2011feature} and also classify tuberculosis \cite{pahar2021tb} and COVID-19 coughs \cite{pahar2020covid, pahar2021covidbreath}. 
We have used the traditional MFCC extraction method considering higher resolution MFCCs along with the velocity (first-order difference, $\Delta$) and acceleration (second-order difference, $\Delta \Delta$) as adding these has shown classifier improvement in the past \cite{azmy2017feature}. 
The ZCR \cite{bachu2010voiced} is the number of times a signal changes its sign within a frame, and indicates the variability present in the signal.
Finally, the kurtosis \cite{decarlo1997meaning} indicates the prevalence of higher amplitudes in the samples of an audio signal. 
These features have been extracted by using the hyperparameters described in Table \ref{table:feat-hyper-parameter} for all cough and non-cough audio events.

\begin{figure}
	\centerline{\includegraphics[width=0.5\textwidth]{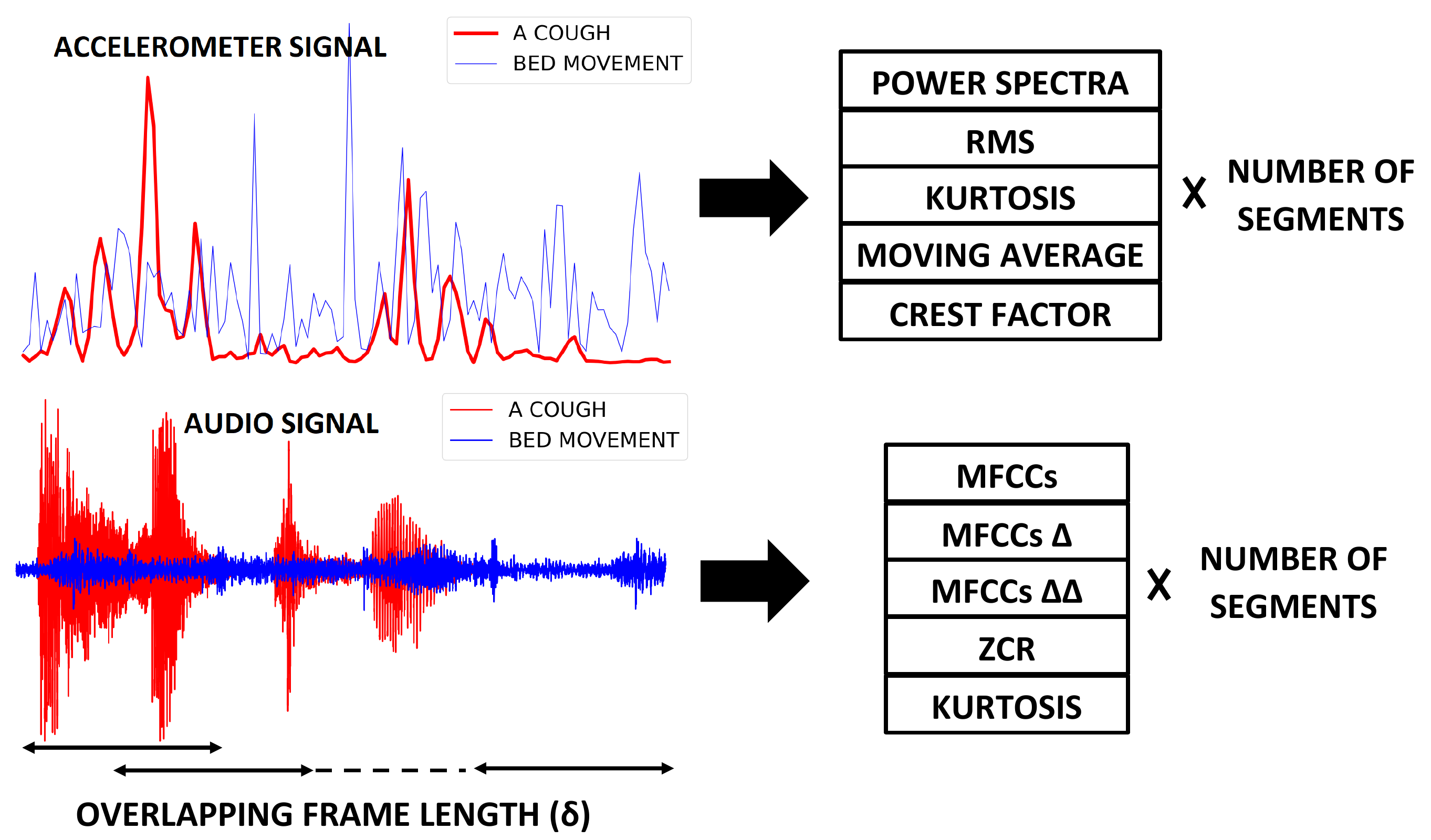}}
	\caption{\textbf{Feature extraction for both accelerometer (top) and audio (bottom) signals:} Both acceleration and audio signals of the events, shown in Figure \ref{fig:cough-Ncough-acc} and \ref{fig:cough-compare}, are split into a fixed number of overlapping frames. The length and number of these frames are $\Psi$ and $C$ for accelerometer signal \& $\mathcal{F}$ and $\mathcal{S}$ for audio signal. For accelerometer measurements, the power spectrum, RMS, kurtosis, moving averages and crest factors of each frame are extracted. For audio signals, the MFCCs, MFCC velocity ($\Delta$), MFCC acceleration ($\Delta \Delta$), ZCR and kurtosis are extracted. For the acceleration signal, this results in a feature matrix with dimensions ($C$, $\frac{\Psi}{2}+5$) while for the audio signal it generates a feature matrix with dimensions ($\mathcal{S}, 3\mathcal{M} + 2$) where $\mathcal{M}$ is the number of MFCCs. }
	\label{fig:feat-extract}
\end{figure}

\subsection{Extraction Process}

The features are extracted in a way that preserves the information regarding the beginning and the end of an event to allow time-domain patterns in the recordings to be discovered while maintaining the fixed input dimensionality, which is expected by the deep neural architectures such as a convolutional neural network (CNN).

For accelerometer signal, the frame length ($\Psi$) and number of segments ($C$) have been used as the feature extraction hyperparameters, shown in Table \ref{table:feat-hyper-parameter}. 
The input feature matrix, fed to the classifiers mentioned in Section \ref{sec:classifier},
has the dimension of ($C$, $\frac{\Psi}{2}+5$) whereas power spectra have $(\frac{\Psi}{2}+1)$ coefficients.

From every event in audio signal, we extract a fixed number of features ($\mathcal{S}$) by distributing the fixed-length analysis frames ($\mathcal{F}$) uniformly over the time-interval of the cough. 
The extracted feature matrix from the audio always has the dimension of ($\mathcal{S}$, $3\mathcal{M} + 2$) for $\mathcal{M}$ number of MFCCs along with $\mathcal{M}$ number of velocity ($\Delta$) and $\mathcal{M}$ number of acceleration ($\Delta \Delta$), as illustrated in Figure \ref{fig:feat-extract}. 

The frame skips are noted as $\delta$ in Figure \ref{fig:feat-extract}. 
We divide the number of samples in an event by the number of segments and take the next positive integer. 
For a 1.2 sec long audio event, the length of frame skip in samples is 
$\ceil[\bigg]{ \frac{1.2 \times 22050}{100} } = \ceil[\bigg]{ \frac{26460}{100} } = 265 $ samples, as the audio sampling rate is 22.05 kHz.

Frame length ($\Psi$) used to extract features from acceleration signal is shorter than the frame length ($\mathcal{F}$) used to extract features from audio in this study (Table \ref{table:feat-hyper-parameter}) and also traditionally \cite{takahashi2016acoustic}. 
This is because the accelerometer in the smartphone has a lower sampling rate of 100 Hz than the microphone {22.05 kHz} (both shown in Figure \ref{fig:recorder-position}) and
longer frames lead to deteriorated performance as the signal properties can no longer be assumed to be stationary \cite{joder2009temporal}.

In contrast with the more conventionally applied fixed, non-overlapping frame rates, this way of extracting features ensures that the entire event is captured within a fixed number of frames, allowing especially the CNN to discover more useful temporal patterns and provide better classification performance. 
This particular method of feature extraction has also shown promising result in classifying COVID-19 coughs, breath and speech \cite{pahar2020covid, pahar2021covidbreath}. 


\begin{table*}[h]
	\footnotesize
	\renewcommand\arraystretch{1.4}
	\setlength\minrowclearance{1.0pt}
	\caption{\textbf{Feature extraction hyperparameters for both accelerometer and audio signals.} For accelerometer, 16, 32, 64 samples i.e. 160, 320 and 640 msec long frames overlap in such a way that the number of these frames i.e. segments (5 and 10) are the same for all events in our dataset. Similarly for audio signals, MFCCs are varied between 13 and 65 \& frames are varied between 256 samples (11.61 msec) and 4096 samples (185.76 msec) in such a way that the number of these extracted frames are varied between 50 to 150, fixed for4 all events in out dataset. }
	\centering 
	\begin{center}
		\begin{tabular}{ c | c | c  }

			\hline
			\hline
			\multicolumn{3}{c}{\textbf{\uppercase{Feature Extraction hyperparameters}}}\\
			\hline
			\hline
			
			\multicolumn{2}{c|}{\textbf{Accelerometer Hyperparameters}} & \textbf{Values} \\
			
			\hline
			Frame ($\Psi$) & Frame-length in samples, used to extract features & $2^k$ where $k=4, 5, 6$ \\
			
			\hline
			Segments ($C$) & Number of frames extracted from the entire event & 5, 10 \\
			
			\hline
			\hline
			
			\multicolumn{2}{c|}{\textbf{Audio Hyperparameters}} & \textbf{Values} \\
			
			\hline
			MFCC ($\mathcal{M}$) & Number of lower-order MFCCs to keep & $13 \times k$, where $k=1, \cdots, 5$ \\
			
			\hline
			Frame ($\mathcal{F}$) & Frame-length in samples, used to extract features & $2^{k}$ where $k=8, \cdots, 12$ \\
			
			\hline
			Segments ($\mathcal{S}$) & Number of frames extracted from the entire event & $10 \times k$, where $k=5, 7, 10, 12, 15$ \\
			
			\hline
			\hline
			
		\end{tabular}
	\end{center}
	\label{table:feat-hyper-parameter}
\end{table*}

\section{Classifier Training}\label{sec:classifier}

We have trained and evaluated six machine learning classifiers on both audio and accelerometer signal. 
Table \ref{table:class-hyper-parameter} lists the classifier hyperparameters that were optimised during leave-one-out cross-validation. 

First, we establish the baseline results by training and evaluating three shallow classifiers such as logistic regression (LR), support vector machine (SVM) and multilayer perceptron (MLP). 
Then, we improve the cough detection performance by implementing three deep neural network (DNN) classifiers such as CNN, long short-term memory (LSTM) and Resnet50.

LR models have outperformed other more complex classifiers such as classification trees, random forests, SVM in some clinical prediction tasks \cite{christodoulou2019systematic, botha2018detection, le1992ridge}. 
The gradient descent weight regularisation as well as lasso ($l1$ penalty) and ridge ($l2$ penalty) estimators \cite{tsuruoka2009stochastic, yamashita2003interior} were the hyperparameters, listed in Table \ref{table:class-hyper-parameter}, optimised inside the nested cross-validation during training. 
SVM classifiers have also performed well in both detecting \cite{bhateja2019pre, tracey2011cough} and classifying \cite{sharan2017cough} cough events in the past.  
The independent term in kernel functions is the hyperparameter optimised for the SVM classifier. 
An MLP, consisting multiple layers of neurons \cite{taud2018multilayer}, is capable of learning non-linear relationships.
It has produced promising results in discriminating influenza coughs from other coughs \cite{sarangi2016design} in the past. 
MLP has also been applied to classify TB coughs \cite{tracey2011cough, pahar2021tb} and detect coughs in general \cite{liu2014cough, amoh2015deepcough}. 
The penalty ratios, along with the number of neurons are used as the hyperparameters, optimised using the leave-one-out cross-validation process (Figure \ref{fig:loo-K-Fold} and Section \ref{sec:classification}).

\begin{figure}
	\centerline{\includegraphics[width=0.5\textwidth]{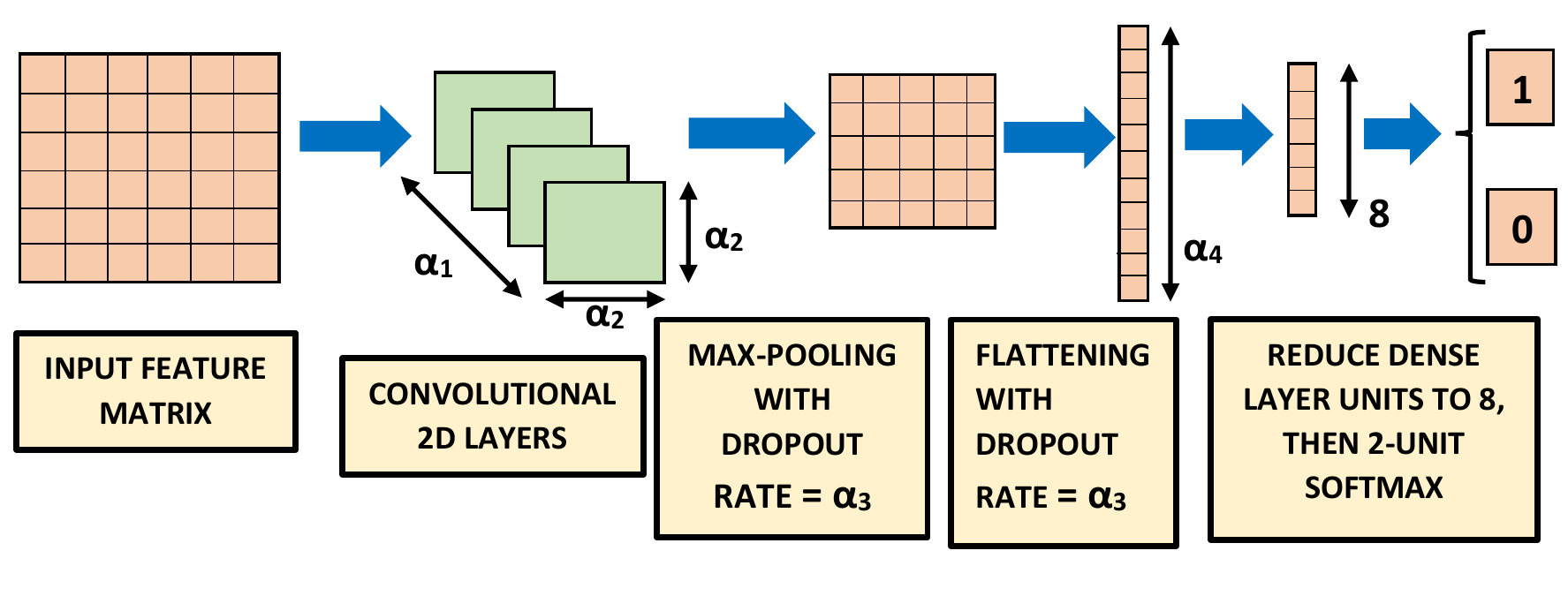}}
	\caption{\textbf{CNN Classifier}, trained and evaluated using leave-one-out cross-validation \cite{sammut2010leave} on 14 patients. The results are shown in Table \ref{table:accelerometer-results} and \ref{table:audio-results} for feature extraction hyperparameters mentioned in Table \ref{table:feat-hyper-parameter}.}
	\label{fig:CNN-fig}
\end{figure}

\begin{figure}
	\centerline{\includegraphics[width=0.5\textwidth]{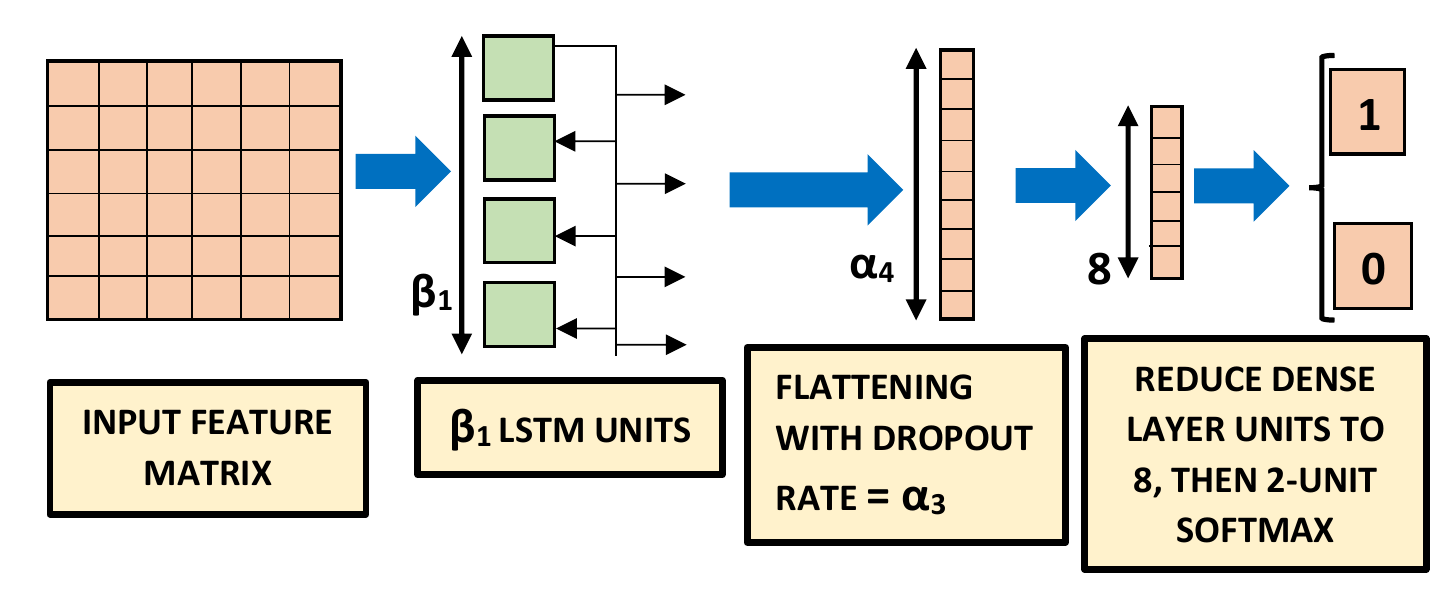}}
	\caption{\textbf{LSTM classifier}, trained and evaluated using leave-one-out cross-validation \cite{sammut2010leave} on 14 patients. The results are shown in Table \ref{table:accelerometer-results} and \ref{table:audio-results} for feature extraction hyperparameters mentioned in Table \ref{table:feat-hyper-parameter}.}
	\label{fig:RNN-fig}
\end{figure}

A CNN is a popular deep neural network architecture, primarily used in image classification \cite{krizhevsky2017imagenet}, such as face recognition \cite{lawrence1997face}. 
It has also performed well in classifying COVID-19 coughs, breath and speech \cite{pahar2020covid, pahar2021covidbreath}. 
The CNN architecture \cite{albawi2017understanding, qi2017comparison}, shown in Figure \ref{fig:CNN-fig}, contains $\alpha_1$ 2D convolutional layers with kernel size $\alpha_2$ and rectified linear units as activation functions. 
A dropout rate $\alpha_3$ has been applied along with max-pooling, followed by 
$\alpha_4$ dense layers with rectified linear units as activation functions, followed by another 8 dense layers, also with rectified linear units as activation functions. 
An LSTM model is a type of recurrent neural network which remembers previously-seen inputs when making its classification decision \cite{hochreiter1997long}. 
It has been successfully used in automatic cough detection \cite{miranda2019comparative, pahar2021deep}, and also in other types of acoustic event detection \cite{marchi2015non, amoh2016deep} including COVID-19 coughs etc. \cite{pahar2020covid, pahar2021covidbreath}. 
The hyperparameters optimised for the LSTM classifier \cite{sherstinsky2020fundamentals} are mentioned in Table \ref{table:class-hyper-parameter} and visually explained in Figure \ref{fig:RNN-fig}. 
The LSTM classifier, shown in Figure \ref{fig:RNN-fig}, contains $\beta_1$ LSTM units with rectified linear units as activation functions and a dropout rate $\alpha_3$. 
Then $\alpha_4$ dense layers have been applied with rectified linear units as activation functions, followed by another 8 dense layers also with rectified linear units as activation functions. 
For both CNN and LSTM classifiers, a final softmax function produces one output for a cough event (i.e. 1) and the other for a non-cough event (i.e. 0), shown in Figure \ref{fig:CNN-fig} and \ref{fig:RNN-fig}. 
Features are fed into these two classifiers in batch size of $\xi_1$ for $\xi_2$ number of epochs. 
The 50-layer residual network (Resnet50) architecture (Table 1 of \cite{he2016deep}) we trained and evaluated has a very deep architecture that contains skip layers and has performed even better than existing deep architectures such as VGGNet on image classification tasks on the dataset such as ILSVRC, the CIFAR10 dataset and the COCO object detection dataset \cite{lin2014microsoft}. 
This architecture has also performed the best in detecting COVID-19 signatures in coughs, breaths and speech \cite{pahar2020covid, pahar2021covidbreath}. 
Due to extreme computation load, we have used the default Resnet50 structure mentioned in Table 1 of \cite{he2016deep}.

\begin{table*}[h]
	\footnotesize
	\renewcommand\arraystretch{1.4}
	\setlength\minrowclearance{1.0pt}
	\caption{\textbf{Classifier hyperparameters}, optimised using the leave-one-patient-out cross-validation. } 
	\centering 
	\begin{center}
		\begin{tabular}{ c | c | c  }
			\hline
			\hline
			\textbf{Hyperparameters} & \textbf{Classifier} & \textbf{Range} \\
			\hline
			\hline
			Regularisation strength ($\gamma_1$) & LR, SVM & $10^i$ where, $i=-7,-6,\ldots,6,7$ \\
			\hline
			$l1$ penalty ($\gamma_2$)  & LR  & 0 to 1 in steps of 0.05 \\
			\hline
			$l2$ penalty ($\gamma_3$)  & LR, MLP  & 0 to 1 in steps of 0.05 \\
			\hline
			Kernel coefficient ($\gamma_4$) & SVM & $10^i$ where, $i=-7,-6,\ldots,6,7$ \\
			
			\hline
			No. of neurons ($\gamma_5$) & MLP & 10 to 100 in steps of 10 \\
			
			\hline
			Batch size ($\xi_1$) & CNN \& LSTM & $2^k$ where $k=6, 7, 8$\\
			\hline
			No. of epochs ($\xi_2$) & CNN \& LSTM & 10 to 250 in steps of 20 \\
			\hline
			No. of convolutional filters ($\alpha_1$) & CNN & $3 \times 2^k$ where $k=3, 4, 5$ \\
			\hline
			kernel size ($\alpha_2$) & CNN & 2 and 3 \\
			\hline
			Dropout rate ($\alpha_3$) & CNN \& LSTM & 0.1 to 0.5 in steps of 0.2 \\
			\hline
			Dense layer size ($\alpha_4$) & CNN \& LSTM & $2^k$ where $k=4, 5$ \\
			\hline
			LSTM units ($\beta_1$) & LSTM & $2^k$ where $k=6, 7, 8$ \\
			\hline
			Learning rate ($\beta_2$) & LSTM & $10^k$ where $k=-2,-3,-4$ \\
			\hline
			\hline
		\end{tabular}
	\end{center}
	\label{table:class-hyper-parameter}
\end{table*}

\section{Classification Process}\label{sec:classification}

\subsection{Hyperparameter optimisation}

Hyperparameters for both the classifiers and feature extraction are optimised inside the leave-one-out cross-validation process and are listed in Table \ref{table:feat-hyper-parameter} and \ref{table:class-hyper-parameter}. 
Different phases of an event carries important information and our special way of feature extraction preserves the time-domain information. 
By varying the frame lengths and number of frames to extract, these information was varied. 
The spectral resolution was also varied by varying the number of lower order MFCCs to keep from the audio signal.

\subsection{Cross-validation}

\begin{figure}
	\centerline{\includegraphics[width=0.5\textwidth]{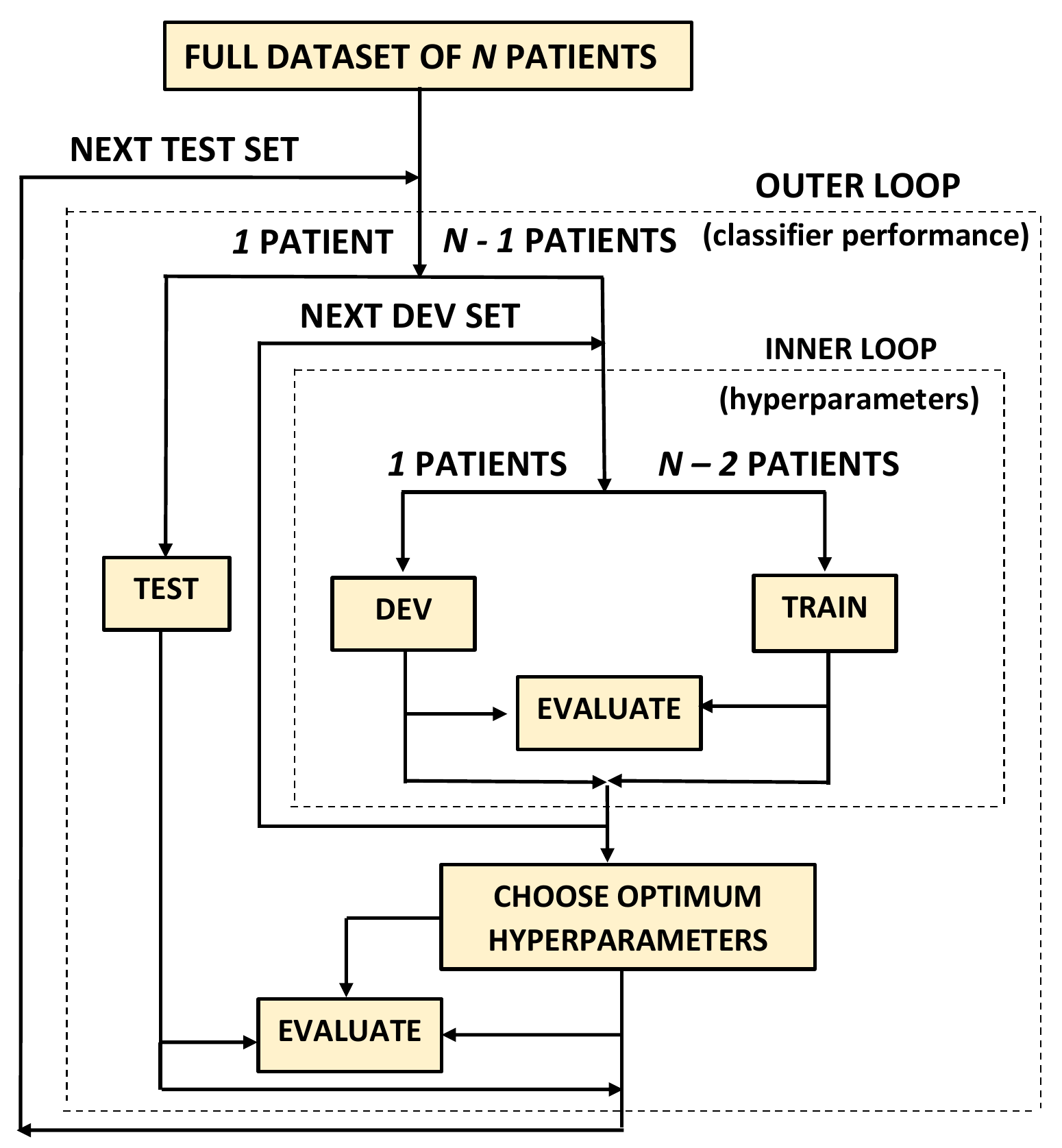}}
	\caption{\textbf{Leave-one-out cross-validation} used to train and evaluate all six classifiers. Here, $N = 14$ (Table \ref{table:Ground-Truth-Dataset-Summary}). The development set (DEV) consisting 1 patient has been used to optimise the hyperparameters while training on the TRAIN set, consisted of 12 patients. The final evaluation of the classifiers in terms of the AUC occurs on the TEST set, consisting 1 patient. }
	\label{fig:loo-K-Fold}
\end{figure}

All six classifiers have been trained and evaluated by using a leave-one-patient-out cross-validation scheme \cite{sammut2010leave}, as explained in Figure \ref{fig:loo-K-Fold}. 
Our dataset contains 14 patients and by using this cross-validation scheme we make the best use of our dataset, as patient's weight, coughing intensity, distance from the microphone can affect the accelerometer and audio signals and we were not allowed to collect those vital information due to the ethical constraints.

The Figure \ref{fig:loo-K-Fold} shows that one patient is left out from 14 patients to be used for later independent testing.
Then another patient is removed from the remaining 13 patients to be used as the development set where the hyperparameters, listed in Table \ref{table:class-hyper-parameter}, are to be optimised. 
AUC has always been the optimisation criterion in this cross-validation. 
This entire procedure is repeated until all patients are used as an independent test sets in the outer loop. 
The final performance is evaluated by calculating and averaging AUC over these outer loops. 
The hyperparameters producing the highest AUC over these outer test sets are noted as the `best hyperparameters' in Table \ref{table:accelerometer-results} and \ref{table:audio-results}. 
Performances produced by each classifier for each set of hyperparameters are noted by `ID' in these tables.

\section{Results}\label{sec:results}

\subsection{Accelerometer-based cough detection}

Table \ref{table:accelerometer-results} lists the performance achieved by all six classifiers for the hyperparameters mentioned in Table \ref{table:feat-hyper-parameter}. 
The results from shallow classifiers are shown in systems C1 to C18 and from deep classifiers are shown in systems C19 to C36 in Table \ref{table:accelerometer-results}. 
These results are the averages over the 14 leave-one-patient-out testing partitions in the outer loop of cross-validation.

The shallow classifiers have provided the baseline classification performance.
Table \ref{table:accelerometer-results} shows that best performance is achieved by an LR classifier with an AUC of 0.8135 along with $\sigma_{AUC}$ of 0.003, specificity of 81.42\%, sensitivity of 81.28\% and an accuracy of 81.35\% (system C4). 
The SVM has produced an AUC of 0.8252 with $\sigma_{AUC}$ of 0.003 and 80.91\% specificity, 84.11\% sensitivity and 82.51\% accuracy while using ten 32 samples long frames (system C10). 
However, the AUC of 0.8587, accuracy of 85.67\%, specificity of 84.47\% and sensitivity of 86.89\% have been achieved from the MLP classifier with 40 neurons and $l2$ penalty ratio of 0.7 using five 64 sample long frames (system C17) and this is the highest AUC achieved by a shallow classifier. 

For the DNN classifiers, the lowest AUC of 0.9243 has been achieved from a CNN classifier in system C19 in Table \ref{table:accelerometer-results}. 
Table \ref{table:accelerometer-results} also shows that the best-performing CNN uses 10 number of 64 samples (640 msec) long frames to achieve an AUC of 0.9499, accuracy of 85.82\%, specificity of 80.91\% and sensitivity of 90.73\% (system C24). 
The optimal LSTM classifier achieves the slightly higher AUC of 0.9572 when using a frame length of 32 samples (320 msec) and 10 of such frames (system C28).
However, the best performance is achieved by the Resnet50 architecture, with an AUC of 0.9888 after 50 epochs from 10 number of 32 samples (320 msec) long frames along with 96.71\% accuracy, 94.09\% specificity and 99.33\% sensitivity (system C34). 

\begin{figure}
	\centerline{\includegraphics[width=0.5\textwidth]{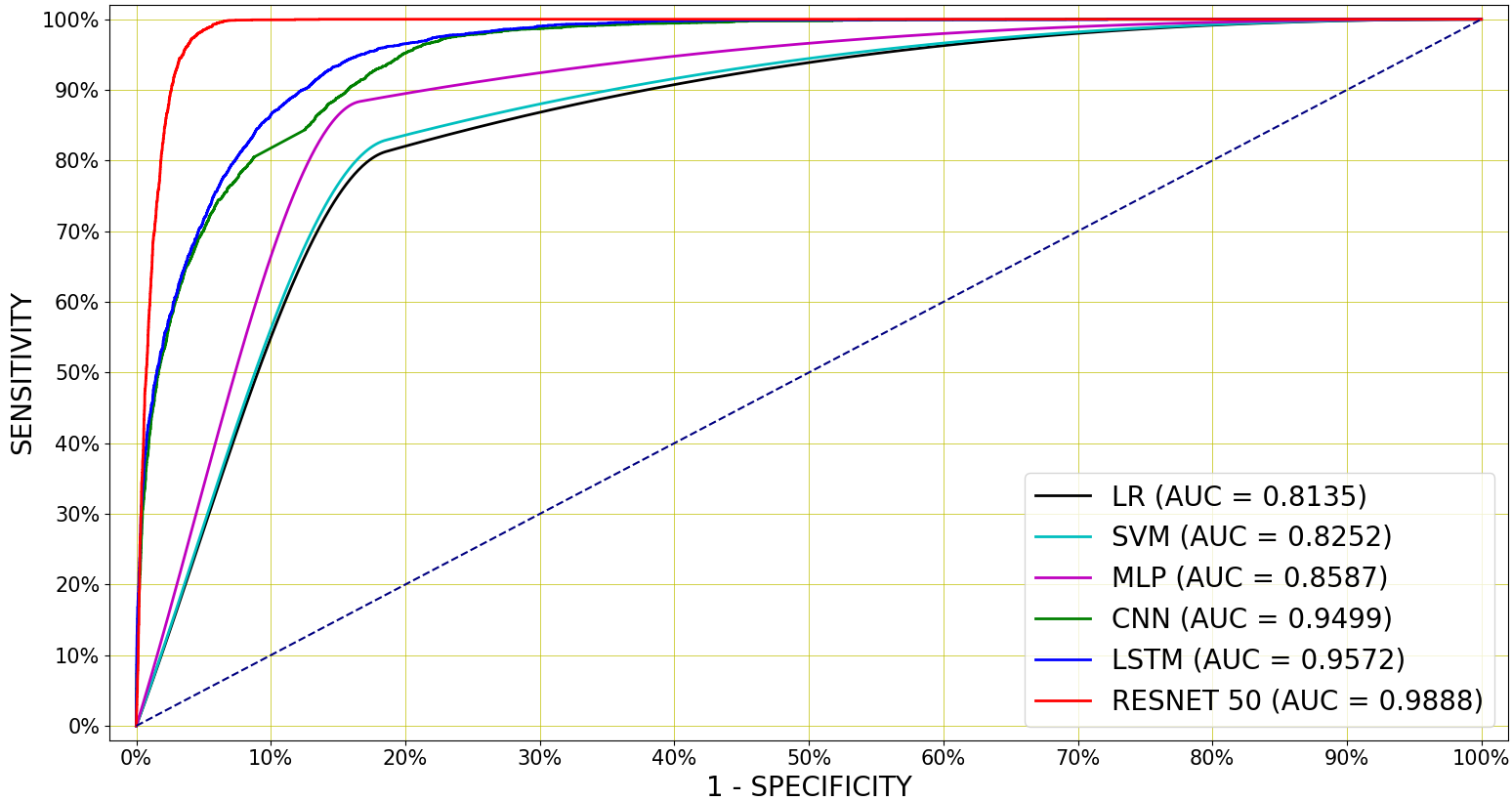}}
	\caption{\textbf{Mean ROC curves for accelerometer-based cough detection}, for the best performing classifiers whose hyperparameters are mentioned in Table \ref{table:class-hyper-parameter}. The best performance has been achieved from a Resnet50, which outperforms the LSTM and CNN over a wide range of operating points and has achieved the AUC of 0.9888 and the accuracy of 96.71\%.}
	\label{fig:mean-ROC-acc}
\end{figure}

Deep architectures have produced a higher AUCs and lower $\sigma_{AUC}$ than the shallow classifiers on accelerometer-based classification task. 
Figure \ref{fig:mean-ROC-acc} shows the mean ROC curves for the optimal LR, SVM, MLP, CNN, LSTM and Resnet50, whose configurations are shown in Table \ref{table:accelerometer-results}, where the means were calculated over the 14 cross-validation folds. 
The Resnet50 classifier is superior to all other classifiers over a wide range of operating points (Figure \ref{fig:mean-ROC-acc}).

\begin{table*}[!h]
	\scriptsize
	\renewcommand\arraystretch{1.4}
	\setlength\minrowclearance{1.0pt}
	\setlength{\tabcolsep}{5pt} 
	\centering
	\caption{\textbf{Accelerometer-based cough detection results.} The values are averaged over 14 cross-validation folds. DNN classifiers have outperformed the shallow classifiers by a wide margin and a Resnet50 produces the highest AUC of 0.9888. } 
	\begin{tabular}{ c | c | c | c | c | c | c | c | c }
		\hline
		\hline
		{\multirow{2}{*}{\textbf{Classifier}}} & {\multirow{2}{*}{\textbf{ID}}} & \textbf{Best Feature} &
		\textbf{Best Classifier Hyperparameters} & \multicolumn{5}{c}{\textbf{Performance}} \\
		
		\cline{5-9}
		
		&  & \textbf{Hyperparameters} & \textbf{(Optimised inside nested cross-validation)} & \textbf{Spec} & \textbf{Sens} & \textbf{Acc} & \textbf{AUC} & \textbf{$\sigma_{AUC}$} \\

		\hline
		\hline
		\multirow{6}{*}{LR} & C1 & $\Psi = 16$, $C = 5$ & $\gamma_1 = 10^{-4}$, $\gamma_2$ = 0.35, $\gamma_3$ = 0.65 & 80.41\% & 80.28\% & 80.35\% & 0.8055 & 0.003 \\
		\cline{2-9}
		& C2 & $\Psi = 16$, $C = 10$ & $\gamma_1 = 10^{-2}$, $\gamma_2$ = 0.55, $\gamma_3$ = 0.45 & 80.25\% & 80.08\% & 80.16\% & 0.8058 & 0.003 \\
		\cline{2-9}
		& C3 & $\Psi = 32$, $C = 5$ & $\gamma_1 = 10^{2}$, $\gamma_2$ = 0.2, $\gamma_3$ = 0.8 & 80.39\% & 80.55\% & 80.47\% & 0.8072 & 0.003 \\
		\cline{2-9}
		& \textit{C4} & \textit{$\Psi = 32$, $C = 10$} & \textit{$\gamma_1 = 10^{-3}$, $\gamma_2$ = 0.4, $\gamma_3$ = 0.6} & \textit{81.42\%} & \textit{81.28\%} & \textit{81.35\%} & \textit{0.8135} & \textit{0.003} \\
		\cline{2-9}
		& C5 & $\Psi = 64$, $C = 5$ & $\gamma_1 = 10^{-1}$, $\gamma_2$ = 0.25, $\gamma_3$ = 0.75 & 80.22\% & 80.41\% & 80.31\% & 0.8119 & 0.003 \\
		\cline{2-9}
		& C6 & $\Psi = 64$, $C = 10$ & $\gamma_1 = 10^{-2}$, $\gamma_2$ = 0.75, $\gamma_3$ = 0.25 & 80.16\% & 80.32\% & 80.24\% & 0.8124 & 0.003 \\

		\hline
		\multirow{6}{*}{SVM} & C7 & $\Psi = 16$, $C = 5$ & $\gamma_1 = 10^{3}$, $\gamma_4 = 10^{-3}$ & 80.71\% & 82.91\% & 81.81\% & 0.8202 & 0.003 \\
		\cline{2-9}
		& C8 & $\Psi = 16$, $C = 10$ & $\gamma_1 = 10^{-2}$, $\gamma_4 = 10^{2}$ & 80.22\% & 82.94\% & 81.58\% & 0.8248 & 0.003 \\
		\cline{2-9}
		& C9 & $\Psi = 32$, $C = 5$ & $\gamma_1 = 10^{3}$, $\gamma_4 = 10^{-2}$ & 80.41\% & 82.97\% & 81.69\% & 0.8212 & 0.003 \\
		\cline{2-9}
		& \textit{C10} & \textit{$\Psi = 32$, $C = 10$} & \textit{$\gamma_1 = 10^{-1}$, $\gamma_4 = 10^{-1}$} & \textit{80.91\%} & \textit{84.11\%} & \textit{82.51\%} & \textit{0.8252} & \textit{0.003} \\
		\cline{2-9}
		& C11 & $\Psi = 64$, $C = 5$ & $\gamma_1 = 10^{-4}$, $\gamma_4 = 10^{-3}$ & 80.28\% & 84.35\% & 82.31\% & 0.8245 & 0.003 \\
		\cline{2-9}
		& C12 & $\Psi = 64$, $C = 10$ & $\gamma_1 = 10^{2}$, $\gamma_4 = 10^{-4}$ & 80.55\% & 82.78\% & 81.68\% & 0.8251 & 0.003 \\

		\hline
		\multirow{6}{*}{MLP} & C13 & $\Psi = 16$, $C = 5$ & $\gamma_3 = 0.55$, $\gamma_5 = 30$ & 82.37\% & 86.95\% & 84.68\% & 0.8507 & 0.003 \\
		\cline{2-9}
		& C14 & $\Psi = 16$, $C = 10$ & $\gamma_3 = 0.45$, $\gamma_5 = 50$ & 83.24\% & 87.08\% & 85.16\% & 0.8558 & 0.003 \\
		\cline{2-9}
		& C15 & $\Psi = 32$, $C = 5$ & $\gamma_3 = 0.35$, $\gamma_5 = 70$ & 83.55\% & 87.41\% & 85.47\% & 0.8552 & 0.003 \\
		\cline{2-9}
		& C16 & $\Psi = 32$, $C = 10$ & $\gamma_3 = 0.4$, $\gamma_5 = 20$ & 82.18\% & 86.05\% & 84.12\% & 0.8424 & 0.003 \\
		\cline{2-9}
		& \textit{C17} & \textit{$\Psi = 64$, $C = 5$} & \textit{$\gamma_3 = 0.7$, $\gamma_5 = 40$} & \textit{84.47\%} & \textit{86.89\%} & \textit{85.67\%} & \textit{0.8587} & \textit{0.003} \\
		\cline{2-9}
		& C18 & $\Psi = 64$, $C = 10$ & $\gamma_3 = 0.35$, $\gamma_5 = 30$ & 83.45\% & 86.84\% & 84.64\% & 0.8499 & 0.003 \\

		\hline
		\multirow{6}{*}{CNN} & C19 & $\Psi = 16$, $C = 5$ & $\alpha_1$=48, $\alpha_2$=2, $\alpha_3$=0.1, $\alpha_4$=32, $\xi_1$=128, $\xi_2$=210 & 83.47\% & 85.62\% & 84.55\% & 0.9243 & 0.002 \\
		\cline{2-9}
		& C20 & $\Psi = 16$, $C = 10$ & $\alpha_1$=24, $\alpha_2$=2, $\alpha_3$=0.3, $\alpha_4$=32, $\xi_1$=256, $\xi_2$=110 & 83.76\% & 87.56\% & 85.66\% & 0.9358 & 0.002 \\
		\cline{2-9}
		& C21 & $\Psi = 32$, $C = 5$ & $\alpha_1$=96, $\alpha_2$=2, $\alpha_3$=0.3, $\alpha_4$=32, $\xi_1$=128, $\xi_2$=150 & 76.98\% & 91.96\% & 84.47\% & 0.9272 & 0.002 \\
		\cline{2-9}
		& C22 & $\Psi = 32$, $C = 10$ & $\alpha_1$=48, $\alpha_2$=2, $\alpha_3$=0.3, $\alpha_4$=16, $\xi_1$=256, $\xi_2$=110 & 84.09\% & 86.41\% & 85.25\% & 0.9324 & 0.002 \\
		\cline{2-9}
		& C23 & $\Psi = 64$, $C = 5$ & $\alpha_1$=48, $\alpha_2$=2, $\alpha_3$=0.5, $\alpha_4$=32, $\xi_1$=256, $\xi_2$=230 & 85.47\% & 87.15\% & 86.31\% & 0.9339 & 0.002 \\
		\cline{2-9}
		& \textit{C24} & \textit{$\Psi = 64$, $C = 10$} & \textit{$\alpha_1$=96, $\alpha_2$=2, $\alpha_3$=0.3, $\alpha_4$=32, $\xi_1$=128, $\xi_2$=170} & \textit{80.91\%} & \textit{90.73\%} & \textit{85.82\%} & \textit{0.9499} & \textit{0.002} \\

		\hline
		\multirow{6}{*}{LSTM} & C25 & $\Psi = 16$, $C = 5$ & $\beta_1$ = 128, $\beta_2$ = 0.0001, $\alpha_3$ = 0.3, $\alpha_4$ = 32, $\xi_1$ = 256, $\xi_2$ = 210 & 84.34\% & 90.82\% & 87.58\% & 0.9444 & 0.002 \\
		\cline{2-9}
		& C26 & $\Psi = 16$, $C = 10$ & $\beta_1$ = 128, $\beta_2$ = 0.01, $\alpha_3$ = 0.1, $\alpha_4$ = 32, $\xi_1$ = 128, $\xi_2$ = 110 & 85.37\% & 91.27\% & 88.32\% & 0.9504 & 0.002 \\
		\cline{2-9}
		& C27 & $\Psi = 32$, $C = 5$ & $\beta_1$ = 256, $\beta_2$ = 0.001, $\alpha_3$ = 0.3, $\alpha_4$ = 16, $\xi_1$ = 128, $\xi_2$ = 130 & 79.92\% & 94.31\% & 87.11\% & 0.9457 & 0.002 \\
		\cline{2-9}
		& \textit{C28} & \textit{$\Psi = 32$, $C = 10$} & \textit{$\beta_1$ = 128, $\beta_2$ = 0.001, $\alpha_3$ = 0.1, $\alpha_4$ = 32, $\xi_1$ = 256, $\xi_2$ = 150} & \textit{86.41\%} & \textit{92.05\%} & \textit{89.21\%} & \textit{0.9572} & \textit{0.002} \\
		\cline{2-9}
		& C29 & $\Psi = 64$, $C = 5$ & $\beta_1$ = 256, $\beta_2$ = 0.001, $\alpha_3$ = 0.3, $\alpha_4$ = 16, $\xi_1$ = 128, $\xi_2$ = 190 & 84.57\% & 92.79\% & 88.68\% & 0.954 & 0.002 \\
		\cline{2-9}
		& C30 & $\Psi = 64$, $C = 10$ & $\beta_1$ = 128, $\beta_2$ = 0.01, $\alpha_3$ = 0.5, $\alpha_4$ = 32, $\xi_1$ = 256, $\xi_2$ = 230 & 86.21\% & 89.13\% & 87.66\% & 0.9489 & 0.002 \\

		\hline
		\multirow{6}{*}{Resnet50} & C31 & $\Psi = 16$, $C = 5$ & Default Resnet50 (Table 1 in \cite{he2016deep}) & 93.81\% & 97.09\% & 95.43\% & 0.9802 & 0.002 \\
		\cline{2-9}
		& C32 & $\Psi = 16$, $C = 10$ & \dittoclosing & 94.12\% & 98.58\% & 96.35\% & 0.9812 & 0.002 \\
		\cline{2-9}
		& C33 & $\Psi = 32$, $C = 5$ & \dittoclosing & 94.29\% & 98.79\% & 96.54\% & 0.9810 & 0.002 \\
		\cline{2-9}
		& \textit{\textbf{C34}} & \textit{\textbf{$\Psi = 32$, $C = 10$}} & \dittoclosing & \textit{\textbf{94.09\%}} & \textit{\textbf{99.33\%}} & \textit{\textbf{96.71\%}} & \textit{\textbf{0.9888}} & \textit{\textbf{0.002}} \\
		\cline{2-9}
		& C35 & $\Psi = 64$, $C = 5$ & \dittoclosing & 94.71\% & 98.23\% & 96.35\% & 0.9854 & 0.002 \\
		\cline{2-9}
		& C36 & $\Psi = 64$, $C = 10$ & \dittoclosing & 95.07\% & 97.89\% & 96.46\% & 0.9884 & 0.002 \\
		
		\hline
		\hline
		
	\end{tabular}
	\label{table:accelerometer-results}
\end{table*}

\subsection{Audio-based cough detection}

\begin{table*}[!h]
	\scriptsize
	\renewcommand\arraystretch{1.4}
	\setlength\minrowclearance{1.0pt}
	\setlength{\tabcolsep}{5pt} 
	\centering
	\caption{\textbf{Audio-based cough detection results.} The values are averaged over 14 cross-validation folds and the best-three performances of each classifier are shown. All classifiers have performed well in detecting coughs but DNN classifiers have performed particularly well and their performances are very close to each other. } 
	\begin{tabular}{ c | c | c | c | c | c | c | c | c }
		\hline
		\hline
		{\multirow{2}{*}{\textbf{Classifier}}} & {\multirow{2}{*}{\textbf{ID}}} & \textbf{Best Feature} &
		\textbf{Best Classifier Hyperparameters} & \multicolumn{5}{c}{\textbf{Performance}} \\
		
		\cline{5-9}
		
		&  & \textbf{Hyperparameters} & \textbf{(Optimised inside nested cross-validation)} & \textbf{Spec} & \textbf{Sens} & \textbf{Acc} & \textbf{AUC} & \textbf{$\sigma_{AUC}$} \\

		\hline
		\hline
		\multirow{3}{*}{LR} & \textit{D1} & \textit{$\mathcal{M} = 26$, $\mathcal{F} = 512$, $\mathcal{S} = 100$} & \textit{$\gamma_1 = 10^{-3}$, $\gamma_2$ = 0.25, $\gamma_3$ = 0.75} & \textit{87.52\%} & \textit{87.71\%} & \textit{87.61\%} & \textit{0.9129} & \textit{0.003} \\
		\cline{2-9}
		& D2 & $\mathcal{M} = 39$, $\mathcal{F} = 1024$, $\mathcal{S} = 70$ & $\gamma_1 = 10^{2}$, $\gamma_2$ = 0.4, $\gamma_3$ = 0.6 & 87.31\% & 87.41\% & 87.36\% & 0.9358 & 0.003 \\
		\cline{2-9}
		& D3 & $\mathcal{M} = 26$, $\mathcal{F} = 1024$, $\mathcal{S} = 100$ & $\gamma_1= 10^{-4}$, $\gamma_2$ = 0.55, $\gamma_3$ = 0.45 & 87.14\% & 87.28\% & 87.21\% & 0.9272 & 0.003 \\

		\hline
		\multirow{3}{*}{SVM} & \textit{D4} & \textit{$\mathcal{M} = 26$, $\mathcal{F} = 1024$, $\mathcal{S} = 120$} & \textit{$\gamma_1 = 10^{-2}$, $\gamma_4= 10^{3}$} & \textit{86.75\%} & \textit{86.91\%} & \textit{86.83\%} & \textit{0.9066} & \textit{0.003} \\
		\cline{2-9}
		& D5 & $\mathcal{M} = 26$, $\mathcal{F} = 512$, $\mathcal{S} = 100$ & $\gamma_1= 10^{2}$, $\gamma_4= 10^{-3}$ & 86.61\% & 86.68\% & 86.64\% & 0.9058 & 0.003 \\
		\cline{2-9}
		& D6 & $\mathcal{M} = 39$, $\mathcal{F} = 1024$, $\mathcal{S} = 100$ & $\gamma_1= 10^{3}$, $\gamma_4= 10^{-3}$ & 86.40\% & 86.54\% & 86.47\% & 0.9017 & 0.003 \\

		\hline
		\multirow{3}{*}{MLP} & \textit{D7} & \textit{$\mathcal{M} = 39$, $\mathcal{F} = 2048$, $\mathcal{S} = 120$} & \textit{$\gamma_3 = 0.35$, $\gamma_5 = 30$} & \textit{89.47\%} & \textit{90.10\%} & \textit{89.78\%} & \textit{0.9254} & \textit{0.002} \\
		\cline{2-9}
		& D8 & $\mathcal{M} = 26$, $\mathcal{F} = 1024$, $\mathcal{S} = 70$ & $\gamma_3 = 0.4$, $\gamma_5 = 50$ & 89.55\% & 89.76\% & 89.66\% & 0.9214 & 0.003 \\
		\cline{2-9}
		& D9 & $\mathcal{M} = 39$, $\mathcal{F} = 1024$, $\mathcal{S} = 100$ & $\gamma_3 = 0.6$, $\gamma_5 = 40$ & 88.78\% & 89.04\% & 88.91\% & 0.9205 & 0.003 \\
		
		\hline
		\multirow{3}{*}{LSTM} & \textit{D10} & \textit{$\mathcal{M} = 26$, $\mathcal{F} = 1024$, $\mathcal{S} = 70$} & \textit{$\beta_1$ = 128, $\beta_2$ = 0.001, $\alpha_3$ = 0.3, $\alpha_4$ = 32, $\xi_1$ = 256, $\xi_2$ = 210} & \textit{94.57\%} & \textit{96.59\%} & \textit{95.58\%} & \textit{0.9932} & \textit{0.002} \\
		\cline{2-9}
		& D11 & $\mathcal{M} = 39$, $\mathcal{F} = 1024$, $\mathcal{S} = 100$ & $\beta_1$ = 128, $\beta_2$ = 0.001, $\alpha_3$ = 0.3, $\alpha_4$ = 16, $\xi_1$ = 256, $\xi_2$ = 130 & 94.21\% & 94.21\% & 96.43\% & 0.9904 & 0.002 \\
		\cline{2-9}
		& D12 & $\mathcal{M} = 26$, $\mathcal{F} = 2048$, $\mathcal{S} = 120$ & $\beta_1$ = 128, $\beta_2$ = 0.01, $\alpha_3$ = 0.3, $\alpha_4$ = 32, $\xi_1$ = 128, $\xi_2$ = 170 & 93.95\% & 96.25\% & 95.10\% & 0.9857 & 0.002 \\
		
		\hline
		\multirow{3}{*}{CNN} & \textit{D13} & \textit{$\mathcal{M} = 26$, $\mathcal{F} = 1024$, $\mathcal{S} = 100$} & \textit{$\alpha_1$ = 48, $\alpha_2$ = 2, $\alpha_3$ = 0.3, $\alpha_4$ = 32, $\xi_1$ = 256, $\xi_2$ = 90} & \textit{93.24\%} & \textit{97.88\%} & \textit{95.56\%} & \textit{0.9944} & \textit{0.002} \\
		\cline{2-9}
		& D14 & $\mathcal{M} = 13$, $\mathcal{F} = 512$, $\mathcal{S} = 70$ & $\alpha_1$ = 24, $\alpha_2$ = 2, $\alpha_3$ = 0.3, $\alpha_4$ = 16, $\xi_1$ = 256, $\xi_2$ = 170 & 92.18\% & 98.74\% & 95.46\% & 0.9891 & 0.002 \\
		\cline{2-9}
		& D15 & $\mathcal{M} = 39$, $\mathcal{F} = 2048$, $\mathcal{S} = 120$ & $\alpha_1$ = 48, $\alpha_2$ = 2, $\alpha_3$ = 0.1, $\alpha_4$ = 32, $\xi_1$ = 256, $\xi_2$ = 130 & 92.78\% & 97.56\% & 95.17\% & 0.9872 & 0.002 \\
		
		\hline
		\multirow{3}{*}{Resnet50} & \textit{\textbf{D16}} & $\mathcal{M} = 26$, $\mathcal{F} = 1024$, $\mathcal{S} = 100$ & Default Resnet50 (Table 1 in \cite{he2016deep}) & \textit{\textbf{96.74\%}} & \textit{\textbf{99.50\%}} & \textit{\textbf{98.13\%}} & \textit{\textbf{0.9957}} & \textit{\textbf{0.001}} \\
		\cline{2-9}
		& D17 & $\mathcal{M} = 39$, $\mathcal{F} = 1024$, $\mathcal{S} = 70$ & \dittoclosing & 96.55\% & 96.95\% & 96.75\% & 0.9912 & 0.001 \\
		\cline{2-9}
		& D18 & $\mathcal{M} = 39$, $\mathcal{F} = 512$, $\mathcal{S} = 100$ & \dittoclosing & 94.70\% & 96.46\% & 95.58\% & 0.9891 & 0.002 \\
		
		\hline
		\hline
		
	\end{tabular}
	\label{table:audio-results}
\end{table*}

To place the performance of the accelerometer-based cough detection presented in the previous section into perspective, we have performed a matching set of experiments, this time using the audio signals to perform audio-based cough detection.
These experiments are based on precisely the same events as the acceleration experiments, since our corpus contains both audio and acceleration signals for each. 

Table \ref{table:audio-results} shows the best-three configurations for each of the six classifier architectures in Section \ref{sec:classifier}.
Again the results indicate that shallow classifiers (LR, SVM and MLP) achieve good classification scores.

LR achieved the highest AUC of 0.9129 with $\sigma_{AUC}$ of 0.003 when using 26 MFCCs, 512 sample long frames and extracting 100 frames (system D1). 
The system has also generated the specificity of 87.52\%, sensitivity of 87.71\% and an accuracy of 87.61\%. 
The SVM achieved an AUC of 0.9066 with $\sigma_{AUC}$ of 0.003 for 26 MFCCs, 1024 sample long frames and extracting 120 frames (system D4). 
The system has also generated the specificity of 86.75\%, sensitivity of 86.91\% and an accuracy of 86.83\%. 
An MLP has has produced the highest AUC of 0.9254 with $\sigma_{AUC}$ of 0.002 for 39 MFCCs, 2048 sample long frames and extracting 120 frames (system D7). 
The system has also generated the specificity of 89.47\%, sensitivity of 90.10\% and an accuracy of 89.78\%. 
This is the best performance achieved by the shallow classifiers. 

Again, the DNN classifiers have outperformed the shallow classifiers by a large margin. 
The best LSTM classifier has produced the highest AUC of 0.9932 with $\sigma_{AUC}$ of 0.002 for 26 MFCCs, 1024 sample long frames and extracting 70 frames (system D10). 
The system has also generated the specificity of 94.57\%, sensitivity of 96.59\% and an accuracy of 95.58\%. 
The best CNN classifier has produced the highest AUC of 0.9944 with $\sigma_{AUC}$ of 0.002 while features were extracted for 26 MFCCs, 1024 sample long frames and extracting 100 frames (system D13). 
The system has also generated the specificity of 93.24\%, sensitivity of 97.88\% and an accuracy of 95.56\%. 
However, the highest AUC of 0.9957 has been achieved again from a Resnet50 classifier with a $\sigma_{AUC}$ of 0.001 for 26 MFCCs, 1024 sample (i.e. 46.44 msec) long frames and extracting 100 frames from the entire event (system D16). 
This system has also achieved a specificity of 96.74\% and a sensitivity of 99.5\% along with the accuracy of 98.13\%.

\begin{figure}
	\centerline{\includegraphics[width=0.5\textwidth]{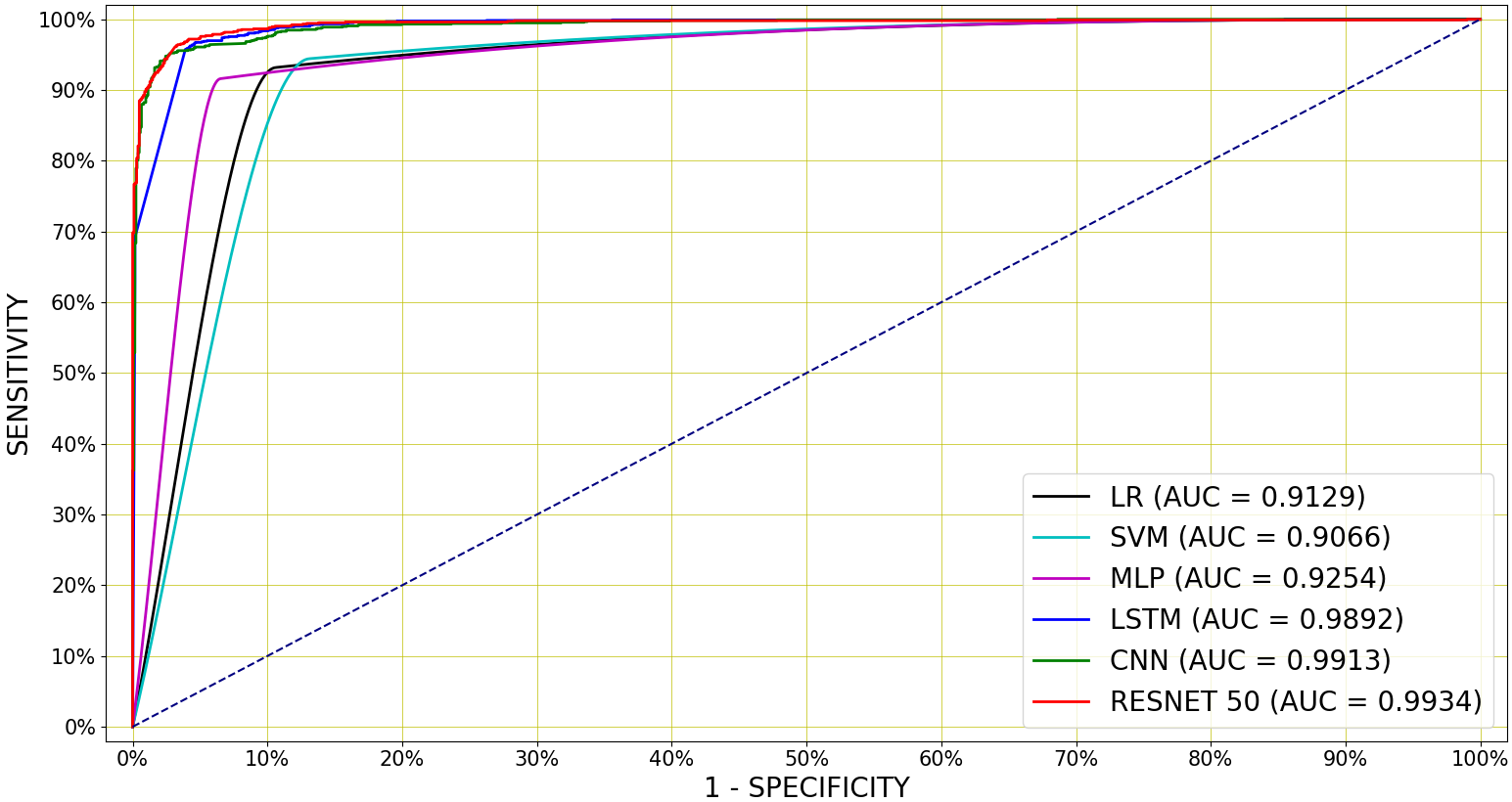}}
	\caption{\textbf{Mean ROC curves for audio-based cough detection}, for the best performing classifiers whose hyperparameters are mentioned in Table \ref{table:class-hyper-parameter}. The best performance has been achieved from a Resnet50 and similar performances have been achieved from a CNN and LSTM. The best Resnet50 produces the AUC of 0.9957 and the accuracy of 98.13\%. }
	\label{fig:mean-ROC-audio}
\end{figure}

Again, deep architectures have produced a higher AUCs and lower $\sigma_{AUC}$ than the shallow classifiers on audio-based classification task. 
These best results for audio-based classification are shown in Figure \ref{fig:mean-ROC-audio}. 
Table \ref{table:audio-results} also indicate that the number of MFCCs has been varied between 13 and 65, although the best performance was achieved by using 26 and 39 MFCCs. 
Using the frame length of 1024 and extracting 100 frames from the events has provided the best performance for most of the classifiers.

\section{Discussion}\label{sec:discussion}

The results shown in Table \ref{table:accelerometer-results} and \ref{table:audio-results} indicate that audio-based cough detection is consistently more accurate than accelerometer-based classification. 
However it is interesting to note that the performances offered by the two alternatives are fairly close. 
In fact, the deep architectures like Resnet50 offer almost equal performance for audio-based and accelerometer-based cough detection. 
It also seems that the CNN and LSTM find it easier to classify cough events based on audio rather than accelerometer signal. 
We postulate that this is due to the limited range of time and frequency information contained in accelerometer data, which is in turn due to the lower accelerometer sampling rate.

For acceleration signals, the extraction of 10 frames each with a length of 640 ms produced the best result.
For the audio, the extraction of 100 frames each with a length of 46.44 ms provided the optimal performance. 
These audio frame lengths are close to those traditionally used for feature extraction in automatic speech recognition.
We also note that the performances of the deep classifiers are consistently better than those offered by the baseline shallow classifiers for both types of signals.
Although the datasets differ, our system also appears to improve on recent work using the accelerometer integrated into a smartwatch~\cite{liaqat2021coughwatch} and detecting cough audio among other audio events such as sneeze, speech, noise etc. \cite{miranda2019comparative}.

\section{Conclusion and Future Work}\label{sec:conclusion}

An automatic non-invasive machine learning based cough detector is able to accurately discriminate between the accelerometer and audio signals due to coughing and due to other movements as captured by a consumer smartphone attached to a patient's bed.


We have trained and evaluated six classifiers including three shallow classifiers such as logistic regression (LR), support vector machine (SVM) and multilayer perceptron (MLP) and three deep neural network (DNN) classifiers such as convolutional neural networks (CNN), long-short-term-memory (LSTM) networks, and a 50-layer residual-based neural network architecture (Resnet50). 
A specially-compiled corpus of manually-annotated acceleration and audio events, including approximately 6000 cough and 68000 non-cough events such as sneezing, throat-clearing and getting in and out of the bed, gathered from 14 adult male patients in a small TB clinic was used to train and evaluate these classifiers by using a leave-one-out cross-validation scheme. 
For accelerometer-based classification, the best system uses a Resnet50 architecture and produces an AUC of 0.9888 as well as a 96.71\% accuracy, 94.09\% specificity and 99.33\% sensitivity while features were extracted from ten 32 sample (320 msec) long frames. 
This demonstrates that it is possible to discriminate between cough events and other non-cough events by using very deep architectures such as a Resnet50; based in signals gathered from an accelerometer that is not attached to the patient's body, but rather to the headboard of the patient's bed. 

We have also compared this accelerometer-based cough detection with audio-based cough detection for the same cough and non-cough events. 
For audio-based classification, the best result has also been achieved from a Resnet50 with the highest AUC of 0.9957. 
This shows that the accelerometer-based cough detection is almost equally accurate as audio-based classification while using very deep architectures such as a Resnet50. 
Shallow classifiers and DNN such as CNN and LSTM however perform better in classifying cough events on audio signals rather than accelerometer signals, as audio signal carries more dynamic and diverse frequency content.

Accelerometer-based detection of cough events has successfully been considered before due to its lower sampling rates and lesser demand of high processing power, however only by using sensors worn by the subjects, which is intrusive and can be inconvenient in some respects. 
This study shows that excellent discrimination is also possible when the sensor is attached to the patient's bed, thus providing a 
less intrusive and cumbersome solution. 
Furthermore, since the use of the acceleration signal avoids the need to gather audio, privacy is inherently protected. 
Therefore, the use of a bed-mounted accelerometer inside an inexpensive consumer smartphone may represent a more convenient, cost-effective and readily accepted method of long-term patient cough monitoring. 

In the future, we will be attempting to optimise some of the Resnet50 metaparameters and fuse both audio and accelerometer signal to achieve higher specificity and accuracy in cough detection. 
We are also in the process of applying the proposed system in an automatic non-invasive cough monitoring system. 
We also note, the manually annotated cough events sometimes contains multiple bursts of cough onsets and we are currently investigating automatic methods that allow such bursts within a cough event to be identified.

\section{Acknowledgements}

We would like to thank the South African Centre for High Performance Computing (CHPC) for providing computational resources on their Lengau cluster for this research.

We thank Mlungisi for his invaluable support in data collection, and gratefully acknowledge the support of Telkom South Africa.

%
\section{Conflict of interest}
The authors declare that they have no conflict of interest.

\bibliographystyle{IEEEbib}      

\bibliography{reference}

\end{document}